\documentclass[twocolumn]{aastex63}

\usepackage{romannum}
\usepackage{amsmath}
\usepackage{multirow}
\usepackage{threeparttable}
\shorttitle{HETDEX: Bright-end Luminosity Functions}
\shortauthors{Zhang et al.}
\begin{document}

\title{First HETDEX Spectroscopic Determinations of Ly$\alpha$ and UV Luminosity Functions at $z=2-3$:\\
Bridging a Gap Between Faint AGN and Bright Galaxies}

\author{Yechi Zhang}
\affiliation{Institute for Cosmic Ray Research, The University of Tokyo, 5-1-5 Kashiwanoha, Kashiwa, Chiba 277-8582, Japan}
\affiliation{Department of Astronomy, Graduate School of Science, the University of Tokyo, 7-3-1 Hongo, Bunkyo, Tokyo 113-0033, Japan}
\correspondingauthor{Yechi Zhang}
\email{yczhang@icrr.u-tokyo.ac.jp}

\author{Masami Ouchi}
\affiliation{National Astronomical Observatory of Japan, 2-21-1 Osawa, Mitaka, Tokyo 181-8588, Japan}
\affiliation{Institute for Cosmic Ray Research, The University of Tokyo, 5-1-5 Kashiwanoha, Kashiwa, Chiba 277-8582, Japan}
\affiliation{Kavli Institute for the Physics and Mathematics of the Universe (Kavli IPMU, WPI), The University of Tokyo, 5-1-5 Kashiwanoha, Kashiwa, Chiba, 277-8583, Japan}

\author{Karl Gebhardt}
\affiliation{Department of Astronomy, The University of Texas at Austin, 2515 Speedway, Stop C1400, Austin, Texas 78712, USA}

\author{Erin Mentuch Cooper}
\affiliation{Department of Astronomy, The University of Texas at Austin, 2515 Speedway, Stop C1400, Austin, Texas 78712, USA}

\author{Chenxu Liu}   
\affiliation{Department of Astronomy, The University of Texas at Austin, 2515 Speedway, Stop C1400, Austin, Texas 78712, USA}

\author{Dustin Davis}   
\affiliation{Department of Astronomy, The University of Texas at Austin, 2515 Speedway, Stop C1400, Austin, Texas 78712, USA}

\author{Donghui Jeong}  
\affiliation{Department of Astronomy and Astrophysics, The Pennsylvania State University, University Park, PA 16802, USA} 
\affiliation{Institute for Gravitation and the Cosmos, The Pennsylvania State University, University Park, PA 16802, USA}

\author{Daniel J. Farrow}  
\affiliation{Max-Planck Institut f\"ur extraterrestrische Physik, Giessenbachstrasse 1, 85748 Garching, Germany}
\affiliation{University Observatory, Fakult\"at f\"ur Physik, Ludwig-Maximilians University Munich, Scheiner Strasse 1, 81679 Munich, Germany}

\author{Steven L. Finkelstein}
\affiliation{Department of Astronomy, The University of Texas at Austin, 2515 Speedway, Stop C1400, Austin, Texas 78712, USA}

\author[0000-0003-1530-8713]{Eric Gawiser}
\affiliation{Department of Physics and Astronomy, Rutgers, The State University of New Jersey, Piscataway, NJ 08854, USA}

\author{Gary J. Hill}
\affiliation{McDonald Observatory, University of Texas at Austin, 2515 Speedway, Stop C1402, Austin, TX 78712, USA}
\affiliation{Department of Astronomy, The University of Texas at Austin, 2515 Speedway, Stop C1400, Austin, Texas 78712, USA}

\author{Yuichi Harikane}
\affiliation{Institute for Cosmic Ray Research, The University of Tokyo, 5-1-5 Kashiwanoha, Kashiwa, Chiba 277-8582, Japan}
\affiliation{Department of Physics and Astronomy, University College London, Gower Street, London WC1E 6BT, UK}

\author{Ryota Kakuma}
\affiliation{Institute for Cosmic Ray Research, The University of Tokyo, 5-1-5 Kashiwanoha, Kashiwa, Chiba 277-8582, Japan}

\author{Viviana Acquaviva}
\affiliation{Physics Department, NYC College of Technology, 300 Jay Street, Brooklyn, NY 11201, USA}
\affiliation{Center for Computational Astrophysics, Flatiron Institute, New York, NY 10010, USA}

\author{Caitlin M. Casey}   
\affiliation{Department of Astronomy, The University of Texas at Austin, 2515 Speedway, Stop C1400, Austin, Texas 78712, USA}

\author{Maximilian Fabricius}
\affiliation{Max-Planck Institut f\"ur extraterrestrische Physik, Giessenbachstrasse 1, 85748 Garching, Germany}
\affiliation{University Observatory, Fakult\"at f\"ur Physik, Ludwig-Maximilians University Munich, Scheiner Strasse 1, 81679 Munich, Germany}

\author{Ulrich Hopp}
\affiliation{Max-Planck Institut f\"ur extraterrestrische Physik, Giessenbachstrasse 1, 85748 Garching, Germany}
\affiliation{University Observatory, Fakult\"at f\"ur Physik, Ludwig-Maximilians University Munich, Scheiner Strasse 1, 81679 Munich, Germany}

\author{Matt J. Jarvis} 
\affiliation{Astrophysics, Department of Physics, Keble Road, Oxford, OX1 3RH, UK}
\affiliation{Department of Physics \& Astronomy, University of the WesternCape, Private Bag X17, Bellville, Cape Town, 7535, South Africa}

\author{Martin Landriau}
\affiliation{Lawrence Berkeley National Laboratory, 1 Cyclotron Road, Berkeley, CA 94720, USA}

\author{Ken Mawatari} 
\affiliation{Institute for Cosmic Ray Research, The University of Tokyo, 5-1-5 Kashiwanoha, Kashiwa, Chiba 277-8582, Japan}
\affiliation{National Astronomical Observatory of Japan, 2-21-1 Osawa, Mitaka, Tokyo 181-8588, Japan}

\author{Shiro Mukae}
\affiliation{Institute for Cosmic Ray Research, The University of Tokyo, 5-1-5 Kashiwanoha, Kashiwa, Chiba 277-8582, Japan}

\author{Yoshiaki Ono} 
\affiliation{Institute for Cosmic Ray Research, The University of Tokyo, 5-1-5 Kashiwanoha, Kashiwa, Chiba 277-8582, Japan}

\author{Nao Sakai}
\affiliation{Institute for Cosmic Ray Research, The University of Tokyo, 5-1-5 Kashiwanoha, Kashiwa, Chiba 277-8582, Japan}

\author{Donald P. Schneider}
\affiliation{Department of Astronomy and Astrophysics, The Pennsylvania State University, University Park, PA 16802, USA}
\affiliation{Institute for Gravitation and the Cosmos, The Pennsylvania State University, University Park, PA 16802, USA}


\begin{abstract}
We present Ly$\alpha$ and ultraviolet-continuum (UV) luminosity functions (LFs) of galaxies and active galactic nuclei (AGN) at $z=2.0-3.5$ determined by the un-targetted optical spectroscopic survey of the Hobby-Eberly Telescope Dark Energy Experiment (HETDEX). We combine deep Subaru imaging with HETDEX spectra resulting in $11.4$~deg$^2$ of fiber-spectra sky coverage, obtaining $18320$ galaxies spectroscopically identified with Ly$\alpha$ emission, $2126$ of which host type 1 AGN showing broad (FWHM$~>1000$~km~s$^{-1}$) Ly$\alpha$ emission lines. We derive the Ly$\alpha$ (UV) LF over 2 orders of magnitude covering bright galaxies and AGN in $\log L_\mathrm{Ly\alpha}/\mathrm{[erg~s^{-1}]}=43.3-45.5$ ($-27<M_\mathrm{UV}<-20$) by the $1/V_\mathrm{max}$ estimator. Our results reveal the bright-end hump of the Ly$\alpha$ LF is composed of type 1 AGN. In conjunction with previous spectroscopic results at the faint end, we measure a slope of the best-fit Schechter function to be $\alpha_\mathrm{Sch}=-1.70^{+0.13}_{-0.14}$, which indicates $\alpha_\mathrm{Sch}$ steepens from $z=2-3$ towards high redshift. Our UV LF agrees well with previous AGN UV LFs, and extends to faint-AGN and bright-galaxy regimes. The number fraction of Ly$\alpha$-emitting objects ($X_\mathrm{LAE}$) increases from $M_\mathrm{UV}^*\sim-21$ to bright magnitude due to the contribution of type 1 AGN, while previous studies claim that $X_\mathrm{Ly\alpha}$ decreases from faint magnitude to $M_\mathrm{UV}^*$, suggesting a valley in the $X_\mathrm{Ly\alpha}-$magnitude relation at $M_\mathrm{UV}^*$. Comparing our UV LF of type 1 AGN at $z=2-3$ with those at $z=0$, we find that the number density of faint ($M_\mathrm{UV}>-21$) type 1 AGN increases from $z\sim2$ to $z\sim0$ as opposed to the evolution of bright ($M_\mathrm{UV}<-21$) type 1 AGN, suggesting AGN downsizing in the rest-frame UV luminosity.

\end{abstract}


\section{Introduction} \label{sec:intro}

At high redshift, Lyman $\alpha$ emitters (LAEs) are a widely studied population of objects that feature strong Lyman $\alpha$ (Ly$\alpha$ $\lambda1216~$\AA) emission lines \citep[e.g.,][]{rhoads00,gronwall07,pentericci09,ouchi20}. Typical LAEs are interpreted as young ($\lesssim 50$~Myr), low-mass ($\lesssim 10^{10}$~M$_{\odot}$) galaxies with high star formation rates (SFR) of $\sim 1-100~$~M$_{\odot}$ yr$^{-1}$ \citep[e.g.,][]{nagao05,gawiser06,finkelstein07,finkelstein08,finkelstein09,ono10b,ono10a,kashikawa12,harikane18}. Such properties make LAEs important tracers for galaxy formation at the low-mass end of the spectrum in the early universe, complementary to the continuum selected Lyman Break Galaxies (LBGs) that are relatively massive.

A key statistical property of LAEs is the luminosity function (LF), which is defined as the number density as a function of luminosity. The LAE LFs and their evolution can provide valuable insights into the evolution of young, star forming (SF) galaxies over the cosmic time. Over the past decades, LAEs have been identified and studied with LFs over the redshift range $z \sim\ 0-9$ using deep narrow band \citep[e.g.,][]{gronwall07,ouchi08,ouchi10,konno16,konno18,tilvi20} and spectroscopic surveys \citep[e.g.,][]{blanc11,cassata11,zheng13,drake17b,herenz19}. These studies have found that at Ly$\alpha$ luminosity $L_\mathrm{Ly\alpha}\lesssim 10^{43}$~erg s$^{-1}$, the Ly$\alpha$ LF of LAEs can be described by the Schechter function \citep{schechter76},
\begin{multline} \label{eq:sch}
    \phi_\mathrm{Sch}(L) \mathrm{dlog}L = \\ \ln10\  \phi^*_\mathrm{Sch}\left(\frac{L}{L^*_\mathrm{Sch}}\right)^{\alpha_\mathrm{Sch}+1}\exp{\left(-\frac{L}{L^*_\mathrm{Sch}}\right)}\mathrm{dlog}L.
\end{multline}
The values of $L^*_\mathrm{Sch}$, $\alpha_\mathrm{Sch}$, and $\phi^*_\mathrm{Sch}$ represent the characteristic luminosity, the faint end slope, and the characteristic number density, respectively. Comparing Ly$\alpha$ LFs at different epochs also shows the redshift evolution of LAEs. From $z \sim 0$ to $z \sim 3$, the number density of LAEs increases rapidly \citep[e.g.,][]{deharveng08}. At $z \sim 3-6$, there is little evolution in the number density of LAEs \citep[e.g.,][]{dawson04,ouchi08}. Beyond $z \sim 6$, the observed number density of LAEs begins to decrease due to the resonant scattering of Ly$\alpha$ photon by the increasing neutral hydrogen (H\Romannum{1}) fraction in the intergalactic medium (IGM) towards the epoch of reionization \citep[e.g.,][]{kashikawa06,hu10,itoh18}, although there is also evidence suggesting no evolution of Ly$\alpha$ LF from $z=5.7-6.5$ \citep[e.g.,][]{malhotra04}.

Despite the efforts of previous studies of the Ly$\alpha$ LFs, several open questions still remain. One open question is the steepness of the faint-end slope of the Ly$\alpha$ LFs that describes the fraction of faint galaxies relative to brighter ones. Theoretical models of hierarchical structure formation predict that low-mass galaxies are more dominant at higher redshift, which results in a steeper faint-end slope. Such a redshift evolution of the faint-end slope has been identified in the UV continuum LF (hereafter UV LF) of LBGs \citep[e.g.,][]{bouwens15,finkelstein15}. Since the dust attenuation of Ly$\alpha$ emission in the interstellar medium (ISM) becomes larger towards fainter UV luminosity (e.g., \citealt{ando06,ouchi08}) and higher redshift (e.g.,\citealt{blanc11,hayes11}), the observed faint-end slope of Ly$\alpha$ LF is expected to be steeper than that of UV LF towards higher redshift. However, the faint-end slope of the Ly$\alpha$ LF is poorly constrained in the previous studies due to the large uncertainty of contamination in photometric LAE samples and the limited spectroscopic LAE samples.

Another open question in relation to the Ly$\alpha$ LF is the shape of the bright end. Several studies have attempted to statistically characterize the bright LAEs with $\log L_\mathrm{Ly\alpha}/\mathrm{[erg~s^{-1}]}\gtrsim 43.5$ based on photometrically selected samples, reaching various conclusions. For example, \citet{konno16} identified an excess in the number density with respect to the Schechter function at the bright end of their Ly$\alpha$ LFs at $z=2.2$. Such an excess was also found in Ly$\alpha$ LFs over other redshift ranges (e.g., \citealt{wold17,zheng17}). \citet{sobral17,sobral18a}, and \citet{matthee17} have shown a similar but less significant bright-end excess at $z=2-3$ and demonstrated that such an excess can be fitted by a power law. At similar redshifts, \citet{spinoso20} found $\sim 14,500$ LAEs with $\log L_\mathrm{Ly\alpha}/\mathrm{[erg~s^{-1}]}>43.3$. The Ly$\alpha$ LFs of their bright LAEs are described by Schechter exponential decays with $\log L^*_\mathrm{Sch}/\mathrm{[erg~s^{-1}]}\sim 44.5-44.8$. Despite these efforts, the precise shape of the Ly$\alpha$ LF, especially at the bright end, is in need of further constraints from spectroscopic studies. 

Along with the shape of the Ly$\alpha$ LF at the bright end, it is also important to understand the nature of the extreme objects that cause the bright-end excess. Although the differential attenuation of Ly$\alpha$ photons in the clumsy ISM may cause the Ly$\alpha$ LFs to have a non-Shechter shape, more evidence attributes the bright-end excess to the existence of Ly$\alpha$-emitting active galactic nuclei (AGN). The AGN activity peaks at $z\sim 2-3$ \citep{hasinger08}, which may result in the contribution of faint AGN to the LAE population. Such a scenario is supported by results based on the spectroscopic follow-ups and multi-wavelength detections of small samples of photometrically-selected, bright LAEs. In particular, \citet{ouchi08} show that their bright LAEs at $z=3.1-3.7$ with $L_\mathrm{Ly\alpha} = 10^{43.3-43.6}$~erg s$^{-1}$ always host AGN. \citet{sobral18b} conduct spectrosopic observations on 21 luminous LAEs at $z \sim 2-3$ and conclude that the AGN fraction increases with Ly$\alpha$ luminosity. Similar trends are also observed through the number fraction of radio and X-ray detected LAEs (e.g., \citealt{matthee17,calhau20}). AGN trace the growth of black holes (BHs) at the center of galaxies and may provide feedback that suppresses star formation in galaxies \citep[e.g.,][]{fabian12,merloni13}. Hence, they are keys to understanding galaxy evolution.

In this paper, we investigate the Ly$\alpha$ and UV LFs of LAEs at $z \sim\ 2-3$ detected by the Hobby-Eberly Telescope Dark Energy Experiment (HETDEX) survey (\citealt{hil08}; Gebhardt et al. in preparation; Hill et al. in preparation). Combining the un-targetted, wide-field integral field spectroscopic (IFS) data of HETDEX and deep ground-based imaging data of Subaru Hyper Supreme Cam (HSC), we explore the low number density regime for both Ly$\alpha$ and UV LFs, where the LAE population consists of both SF galaxies and AGN.

This paper is arranged as follows. Section 2 describes the details of the HETDEX survey and the spectroscopic data included in this paper. Our LAE samples are presented in Section 3. In Section 4, we derive the Ly$\alpha$ LF at $z=2.0-3.5$. We also show the spectroscopic properties and the UV LF of type 1 AGN in our LAE samples. We discuss the evolution of the Ly$\alpha$ LF of LAEs and the UV LF of type 1 AGN in Section 5. Throughout this paper, we use AB magnitudes \citep{oke74} and the cosmological parameters of ($\Omega_\mathrm{m}$, $\Omega_{\Lambda}$, $h$) = (0.3, 0.7, 0.7). 

\section{Observations and Data} \label{sec:data}
\subsection{HETDEX Survey}

HETDEX is an un-targeted integral field spectroscopic survey designed to measure the expansion history of the universe at $z~\sim~1.9-3.5$ by mapping the three-dimensional positions of around 1 million LAEs. The survey started in January 2017, and is scheduled to complete in 2024. On completion, it will cover $\sim$540 deg$^2$ of sky area that is divided into northern (''Spring'') and equatorial (''Fall'') fields. The corresponding survey volume is $\sim11$ comoving Gpc$^3$. The survey is conducted with the Visible Integral-field Replicable Unit Spectrograph 
(\citealt{hil18a}; Hill et al. in preparation),
which is fed by fibers from the prime focus of the upgraded 10m Hobby-Eberly Telescope (HET, \citealt{lwr94,hil18b}; Hill et al. in preparation). VIRUS is a replicated integral field spectrograph \citep{hil14} that consists of 156 identical spectrographs (arrayed as 78 units, each with a pair of spectrographs) fed by 34,944 fibers, each $1.''5$ diameter, projected on sky.  VIRUS has a fixed spectral bandpass of $3500-5500$~\AA\ and resolving power R~$\sim$~800 at $4500$~\AA~ (\citealt{hil18a}; Hill et al. in preparation). The fibers are grouped into 78 integral field units (IFUs, \citealt{kelz14}), each with 448 fibers in a common cable. There is one IFU covering $51''\times51''$ area for each two-channel spectrograph unit. The fibers are illuminated directly at the f/3.65 prime focus of HET and are arrayed within each IFU with a 1$/$3 fill-factor such that an observation requires three exposures with dithers in sky position to fill in the areas of the IFUs. For HETDEX, the exposure time is 3x360 seconds for each pointing set. The IFUs are arrayed in a grid pattern with 100 arcsecond spacing, within the central 18 arcmin diameter of the field of the upgraded HET, and fill this area with $\sim$1$/$4.5 fill factor. A detailed technical description of the HET wide field upgrade and VIRUS and their performance is presented in Hill et al. (in preparation).

The data presented in this paper were obtained as part of the internal HETDEX data release 2 (iHDR2, Gebhardt et al., in preparation). iHDR2 includes 3086 exposure sets taken between January 2017 and June 2020 with between 16 and 71 active IFUs. From these data we select $1862$ exposure sets whose footprints are covered by Subaru/Hyper Suprime-Cam (HSC) imaging data. The total area is $57$~deg$^2$ with spatial filling factors of $\sim 20${\%}, which yields an effective area of $11.4$~deg$^2$, corresponding to $\sim 2.0\times10^8$~comoving Mpc$^3$ for the redshift range of $2.0<z<3.5$. 

\subsection{Subaru HSC Imaging}\label{subsec:hsc}
To provide better measurements on the UV continuua of LAEs at $z=1.9-3.5$, we utilize the $r$-band imaging data taken by Subaru/HSC. The HSC $r$-band filter covers the wavelength range of $5500~\sim~7000$~\AA. For sources at $z=1.9-3.5$, the HSC $r$-band magnitudes serve as a good estimation for the UV continua while the HETDEX spectra can detect the Ly$\alpha$ emission lines. 

The $r$-band imaging data in this study are taken from two surveys, the HSC $r$-band imaging survey for HETDEX (hereafter HETDEX-HSC survey) and Subaru Strategic Program \citep[HSC-SSP;][]{aihara18}. HETDEX-HSC survey has obtained imaging data of a $\sim$250~deg$^2$ area in the Spring field. The observations of the HETDEX-HSC survey were carried out in 2015-2018 (S15A, S17A, S18A; PI: A. Schulze) and 2019-2020 (S19B; PI: S. Mukae) with the total observing time of 3 nights and the seeing sizes of $0''.6-1''.0$. The 5$\sigma$ limiting magnitude in a $3''.0$ diameter aperture is $r=25.1$ mag.

In addition to the HETDEX-HSC survey, we also exploit the $r$-band imaging data in the public data release 2 (PDR2) of HSC-SSP \citep{aihara19}. The HSC-SSP PDR2 includes deep multi-color imaging data of a sky area of $\sim 300$~deg$^2$ taken over the span of 300 nights. The $r$-band imaging data of HSC-SSP PDR2 have average seeing size of $\sim 0''.7$. The 5$\sigma$ limiting magnitude for the $3''.0$ diameter aperture is $r=25.8$ mag and $r=27.7$ mag in the Wide (W) and UltraDeep (UD) layers, respectively. The data reduction and source detection of HETDEX-HSC and HSC-SSP surveys are conducted with \texttt{hscPipe} \citep{bosch18} version 6.7.

\section{Samples} \label{sec:sample}
In this section, we provide details of how we construct LAE samples with iHDR2. Although iHDR2 includes a curated emission line catalog (hereafter HETDEX emission line catalog) that is based on the blind search for detections from all spectral and spatial elements (Gebhardt et al., in preparation), the HETDEX emission line catalog fails to recover some of the  previously identified type 1 AGN with broad emission lines. This is because the HETDEX emission line catalog is optimized for typical star-forming galaxies with narrow emission lines. Given the large HETDEX dataset, current attempts to include broad emission line candidates remain challenging, as this could introduce artifacts such as continuum between two close absorption lines and humps caused by calibration issues. With the challenge of selecting broad-line LAEs from the HETDEX emission line catalog, we construct a new emission line catalog based on the iHDR2 reduced fiber spectra and the deep HSC $r$-band imaging data (hereafter HSC-detected catalog). Our detection algorithm performs emission line detection in a variety of wavelength bins at the positions of continuum sources in the HSC $r$-band imaging data. Because the spatial positions are determined, the algorithm is able to detect broad emission line candidates while limiting the number of artifacts. In Section \ref{subsec:emission_lines}, we describe the HETDEX emission line catalog and the HSC-detected catalog. We combine the HETDEX emission line catalog and the HSC-detected catalog, making an emission line catalog with signal-to-noise ratio (S/N) $> 5.5$ that does not miss broad emission lines. In Section \ref{subsec:laesel}, we construct the LAE sample from the combined catalog. Section \ref{subsec:followup} describes our spectroscopic follow-ups on LAE candidates selected in Section \ref{subsec:laesel}.

\subsection{Emission Line Catalogs}\label{subsec:emission_lines}
\subsubsection{HETDEX Emission Line Catalog}\label{subsec:het_laes}

The HETDEX emission line catalog (internally v2.1.1) is constructed with an automatic detection pipeline developed by the HETDEX collaboration. Details of the pipeline are introduced in Gebhardt et al. (in preparation). In general, the following three steps are involved. i) The positions of emission line candidates are determined based on a grid search in all spatial and spectral elements of the reduced HETDEX fiber spectra with bin sizes of $0.''5$ in the spatial direction and 4\AA\ in the wavelength direction. ii) At the positions of detections, 1D spectra are extracted, and emission line fits are conducted to measure the central wavelengths, fluxes, and linewidths of emission line candidates.  iii) Emission line candidates are further screened based on the $\chi^2$ of the fitting results, S/N, and linewidths.

From the HETDEX emission line catalog, we obtain emission line candidates with S/N $> 5.5$ and wavelength of 3666~\AA $< \lambda <$ 5490~\AA. The S/N $> 5.5$ cut allow us to obtain a relatively clean emission line sample with limited false detections. We apply the wavelength cut to keep consistency with our HSC-detected sources (Section \ref{subsec:hsc_laes}). We also require the emission line candidates to locate within the areas covered by HSC $r$-band imaging data. We measure the rest-frame UV continua of emission line candidates using the HSC $r$-band imaging data. We cross-match the emission line candidates with the HSC $r$-band detected sources within $2''$ radii, and use the $3''$ diameter aperture magnitudes of the $r$-band detected sources as the continua flux density. If there are no $r$-band detected counterparts, we use the $5\sigma$ limiting magnitude (Table \ref{tab:lae}) as the upper limit of the continua flux density.  

To check the possible contamination from false detections such as cosmic rays, sky residuals, and bad pixels, we randomly select $412$ emission line candidates and conduct visual classification. We find that $17/412$ ($4\%$) emission line candidates are false detections. This indicates a negligible fraction of false detections in our HETDEX emission line catalog. We obtain $138314$ emission line candidates from the HETDEX emission line catalog. 

\subsubsection{HSC-Detected Catalog}\label{subsec:hsc_laes}
We construct our HSC-detected emission line catalog based on the $r$-band imaging data of the HETDEX-HSC survey and the HSC-SSP survey. From the $r$-band detected source catalog, we select isolated or cleanly deblended sources. We then require that none of the central $3 \times 3$ pixels are saturated, and none of the central $3 \times 3$ pixels are severely affected by very bright neighboring sources. We also remove objects with flags indicating failed centroid position measurements. The selection criteria that we use for this purpose are listed in Table \ref{tab:hsc_flags}. We then limit the catalog to the $2972750$ sources whose S/N is larger than $5$ based on their $3''$ diameter aperture magnitudes.

\begin{table*}[ht!]
\centering
\caption{Flags for the HSC $r$-band source catalog} \label{tab:hsc_flags}
\begin{tabular}{ccc}
\hline
\hline
Parameter & Value & Notes \\
\hline
r\_pixelflags\_edge & False & Source is outside usable exposure region \\
r\_pixelflags\_interpolatedcenter & False & Interpolated pixel in the source center \\
r\_pixelflags\_saturatedcenter & False & Saturated pixel in the source center \\
r\_pixelflags\_crcenter & False & Cosmic ray in the source center \\
r\_pixelflags\_bad & False & Bad pixel in the source footprint \\
r\_pixelflags\_bright\_objectcenter & False & Source center is close to BRIGHT\_OBJECT pixels \\
r\_pixelflags\_bright\_object & False & Source footprint includes BRIGHT\_OBJECT pixels \\
r\_pixelflags & False & General failure flag \\
detect\_istractinner & True & True if source is in the inner region of a coadd tract \\
detect\_ispatchinner & True & True if source is in the inner region of a coadd patch \\
\hline
\end{tabular}
\end{table*}

We extract 1D source spectra $F_{\mathrm{source}}(\lambda)$ at the positions of the objects in the $r$-band source catalog from the reduced HETDEX fiber spectra. We assume that the 2D distribution of light on the focal plane for point-sources is described by the Moffat PSF \citep[e.g.][]{moffat69,trujillo01}, and calculate the PSF value $p_i$ at the position of fiber $i$. We sum up the fiber spectra within a radius of $2.''5$ around the positions of the objects in the $r$-band source catalog using the equation:
\begin{equation}
F_{\mathrm{source}}(\lambda) = \frac{\sum_{i=1}^n F_i(\lambda)p_i}{\sum_{i=1}^n p_i^2},
\end{equation}
where $F_{i}(\lambda)$ is the spectrum of fiber $i$. Since the noise levels of fiber spectra at the edge of the HETDEX spectral range is $>3$ times higher than the median level, we only use the data in the spectral range of 3666~\AA~$<~\lambda <$~5490~\AA. This wavelength range corresponds to the Ly$\alpha$ redshift of 2.0 $< z <$ 3.5. 

We perform emission line detection and flux measurement on the source spectra. We first subtract continua from the source spectra with the median filter and sigma clipping method. To include the broad emission lines, we then repeatedly scan and measure the S/N of the source spectra within various wavelenth bins (6, 10, 18, 34, and 110 \AA). We require the emission lines to have S/N $> 5.5$ in any of the wavelength bins. We fit single Gaussian profiles to measure the central wavelengths and FWHMs of the emission lines. Since the shapes of Ly$\alpha$ emission lines are sometimes asymmetric or have double peaks \citep[e.g.][]{dijkstra06,matthee18}, we additionally fit double Gaussian profiles. For flux measurements, the choice between the single Gaussian and the double Gaussian models are determined by visual inspections.

To remove false detections such as cosmic rays, sky residuals, and bad pixels, we first apply the machine learning (ML) classifier \citep[][Sakai et al., in preparation]{sakai21} that takes the 2D spectrum of each emission line candidate as input and returns a score from 0 (false detections) to 1 (real detections). We test the reliability of our ML classifier by visually classifying $14276$ randomly selected detections. Among the $14276$ detections, $1725$($12551$) are classified as real(false) detections. Comparing our visual classification with the scores given by the ML classifier, we find that a score higher than $0.1$ effectively remove $5762/12551$($46\%$) false detections while recovering $1624/1725$($94\%$) real detections. We thus require a score higher than $0.1$. This initial screening process yields to $385493$ emission line candidates from the HSC-detected catalog. We conduct visual classification to remove the remaining false detections in the HSC-detected catalog after selecting LAE candidates in Section \ref{subsec:laesel}.

\subsection{LAE Selection}\label{subsec:laesel}
It is challenging to distinguish Ly$\alpha$ from {\sc [Oii]} $\lambda3727$~\AA \ lines with the HETDEX spectral data alone for three reasons. First, the spectral resolution of VIRUS is not high enough to distinguish between the asymmetric Ly$\alpha$ and the blended {\sc [Oii]} doublet based on the skewness of their emission line profiles (Hill et al. in preparation; Gebhardt et al. in preparation). Second, many LAEs at $2.06~\leq~z~\leq~3.5$ and {\sc [Oii]} emitters at $0.1~\leq~z~\leq~0.5$ appear as single-line emitters in the HETDEX spectra whose range is $3500-5500$~\AA. Finally, the relatively low sensitivity makes it difficult to precisely measure the continuum of sources within the spectrum alone to select LAEs with the rest-frame equivalent width (EW$_0$) cut technique \citep[e.g.][]{gronwall07,ouchi08,konno16} that is widely used.

To resolve this problem, we identify Ly$\alpha$ emission lines from the final emission line catalog based on the EW$_0$ calculated from the flux measured from HETDEX spectra and continuum measured from HSC $r$-band imaging data. Our goal is to select Ly$\alpha$ emission lines from both SF galaxies and AGN.
We first require the EW$_0$ of the emission line to be greater than 20~\AA, assuming Ly$\alpha$ redshifts. This Ly$\alpha$ EW$_0$ cut is similar to previous LAE studies \citep[e.g.][]{gronwall07,konno16}. We estimate continuum flux densities at the wavelength of Ly$\alpha$ from the HSC $r$-band magnitudes with the assumption of a flat UV continuum (i.e., $F_\nu$ = const). This is a typical UV spectrum assumed in studies on high-z galaxies including LAEs (e.g., \citealt{ouchi08,konno16,konno18}). At $z = 2-3.5$, the $r$-band central wavelength corresponds to the rest-frame $1400-2000$~\AA. If we assume a UV slope of $-0.5$ to $0.5$, the resulting UV flux changes by only $20\%$. 

We notice that with this Ly$\alpha$ EW$_0$ cut, there still remain a consistent number of foreground contaminants. Most of these contaminants are the {\sc [Oii]} ({\sc Civ} $\lambda 1549$~\AA, {\sc Ciii}] $\lambda 1909$~\AA, and Mg{\sc ii} $\lambda 2798$~\AA) emission lines from low-z SF galaxies (AGN). Because the sources of contamination are different for SF galaxies and AGN, it is difficult to select Ly$\alpha$ emission lines from SF galaxies and AGN with one set of criteria. To resolve this problem, We use the FWHM of the emission line to separate SF galaxies and type 1 AGN (e.g., \citealt{netzer90}). We divide our EW$_0>20$~\AA\ emission line sample into two subsamples, narrow line (NL, FWHM $\leq 1000$ km s$^{-1}$) and broad line (BL, FWHM $> 1000$ km s$^{-1}$) subsamples, that mainly contain SF galaxies and type 1 AGN, respectively. We select NL-(BL-) LAEs from NL (BL) subsample, combining them to make our combined LAE (C-LAE) sample. Section \ref{subsec:nllae} and \ref{subsec:bllae} discuss NL- and BL- LAE selection respectively. We summarize our LAE sample in Section \ref{subsec:laesum}.

\subsubsection{NL-LAE Subsample} \label{subsec:nllae}
\begin{figure*}[ht!]
\begin{center}
\includegraphics[scale=0.55]{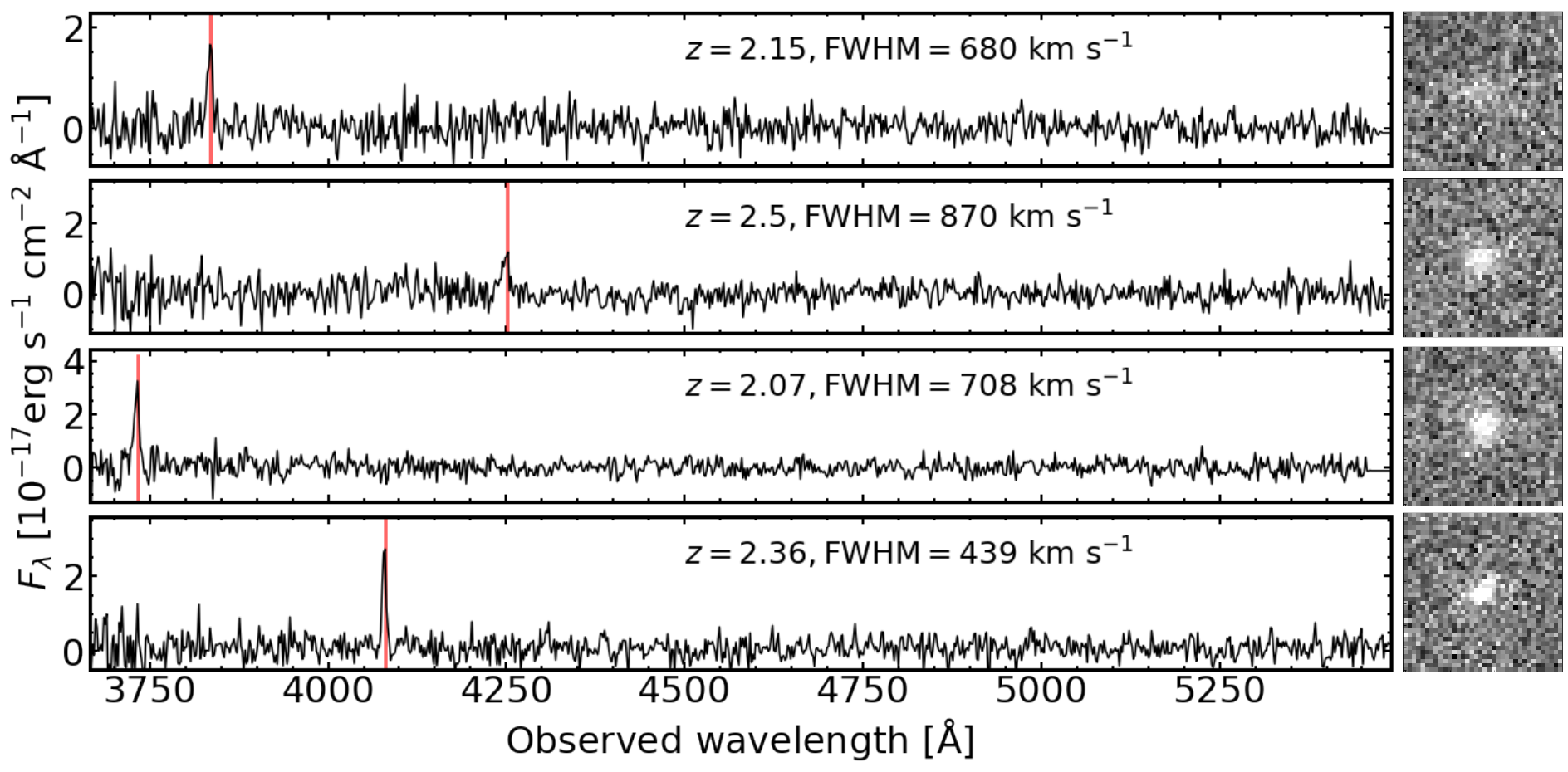}
\end{center}
\caption{Examples of NL-LAEs. Left panels show the HETDEX spectra (black lines) of NL-LAEs. The redshifts and FWHMs are indicated at the top of each spectrum. The red solid lines represent the wavelengths of Ly$\alpha$ emission lines. Right panels denote the $8''\times8''$ $r$-band images from the HETDEX-HSC survey. North is up, and east is to the left.\label{fig:nl_laes}}
\end{figure*}

We select NL-LAEs from the NL emission line catalog with EW$_0 > 20$ \AA\ and FWHM$\leq 1000$ km s$^{-1}$. For the typical SF galaxy, the most prominent emission lines within the HETDEX spectral range of $3500-5500$ \AA\ could be Ly$\alpha$, {\sc [Oii]}, H$\beta$ $\lambda$4863 \AA, and {\sc [Oiii]} $\lambda$4960~\AA, $\lambda$5007~\AA. Assuming the single emission line to be H$\beta$ at $z=0-0.13$ ({\sc [Oiii]} at $z=0-0.10$), the HETDEX spectra would cover other strong emission lines such as {\sc [Oii]} (H$\beta$) on the blue (red) side of the spectrum. Because other emission lines of LAEs at $z=1.9-3.5$ ({\sc [Oii]} emitters at $z=0.13-0.48$) lie beyond the HETDEX wavelength coverage, LAEs ({\sc [Oii]} emitters) would appear as single-line emitters in the HETDEX spectra. Due to this reason, LAEs at $z=2.1-3.5$ and {\sc [Oii]} emitters at $z=0-0.5$ are indistinguishable in the HETDEX spectra for typical SF galaxy. To isolate LAEs from {\sc [Oii]} emitters, we apply the Bayesian statistical method \citep{leung17a}, as modified and implemented by \citet{farrow21} and Davis et al (in preparation). The Bayesian statistical method calculates the probability $P_\mathrm{{LAE}}$ ($P_\mathrm{{[O\Romannum{2}]}}$) that a given source is an LAE ({\sc [Oii]} emitter) based on the $\mathrm{EW_0}$, the luminosity, and the wavelength of the source. The probability ratio $P_\mathrm{{LAE}}/P_\mathrm{[OII]}$ serves as the LAE selection criterion. We test various possible LAE selection criteria ($P_\mathrm{{LAE}}/P_\mathrm{[OII]} > 1,2,5,10,20$) with the HETDEX iHDR2 data, where $37$ $(10)$ spectroscopically and photometrically identified LAEs (foreground contaminants) are included in our emission line catalog \citep[e.g.][]{kriek15,laigle16,tasca17} with EW$_0 > 20$ \AA. For each possible LAE selection criterion, we make test samples of both confirmed LAEs and foreground contaminants that meet the criterion. We calculate fractions of confirmed LAEs (foreground contaminants) in the test samples to the total confirmed LAEs (test samples) that correspond to a sample completeness (contamination rate). We apply the criterion of $P_\mathrm{{LAE}}/P_\mathrm{[OII]} > 1$ that provides a reasonably high completeness of $76\%$ (=28/37) and a reasonably low contamination rate of $13\%$ (=4/32). The number of objects in our NL-LAE subample is $16194$. Figure \ref{fig:nl_laes} shows four examples of our NL-LAEs.

\begin{deluxetable*}{cccccccc}[ht!]
\tablecaption{Summary of LAE samples \label{tab:lae}} 
\tablecolumns{8}
\tablewidth{0pt}
\tablehead{
\colhead{Field} &
\colhead{(R.A., Decl.)} &
\colhead{Area\tablenotemark{a}} &
\colhead{$N_\mathrm{NL-LAE}$} & 
\colhead{$N_\mathrm{BL-LAE}$} &
\colhead{$N_\mathrm{C-LAE}$} &
\colhead{Imaging data} &
\colhead{$r_{5\sigma}$\tablenotemark{b}} \\
\colhead{} & \colhead{(deg)} & \colhead{(arcmin$^2$)} &
\colhead{} & \colhead{} & \colhead{} & \colhead{} & \colhead{(mag)}
}
\startdata
dex-Spring & ($205.0$, $51.3$) & $29470$ & $11966$ & $1532$ & $13498$ & HET-HSC & $25.1$ \\ 
dex-Fall & ($21.5$, $0.0$) & $11542$ & $4129$ & $584$ & $4713$ & HSC-SSP(W) & $25.8$ \\ 
COSMOS & ($150.2$, $2.3$) & $133$ & $99$ & $10$ & $109$ & HSC-SSP(UD) & 27.7 \\ 
\hline
Total & & $41145$ & $16194$ & $2126$ & $18320$ & & \\ 
\enddata
\tablenotetext{a}{Effective survey area covered by HETDEX fibers.}
\tablenotetext{b}{$5\sigma$ limiting magnitude of imaging data for a $3''$ diameter aperture.}
\end{deluxetable*}

\subsubsection{BL-LAE Subsample} \label{subsec:bllae}
\begin{figure*}[ht!]
\begin{center}
\includegraphics[scale=0.55]{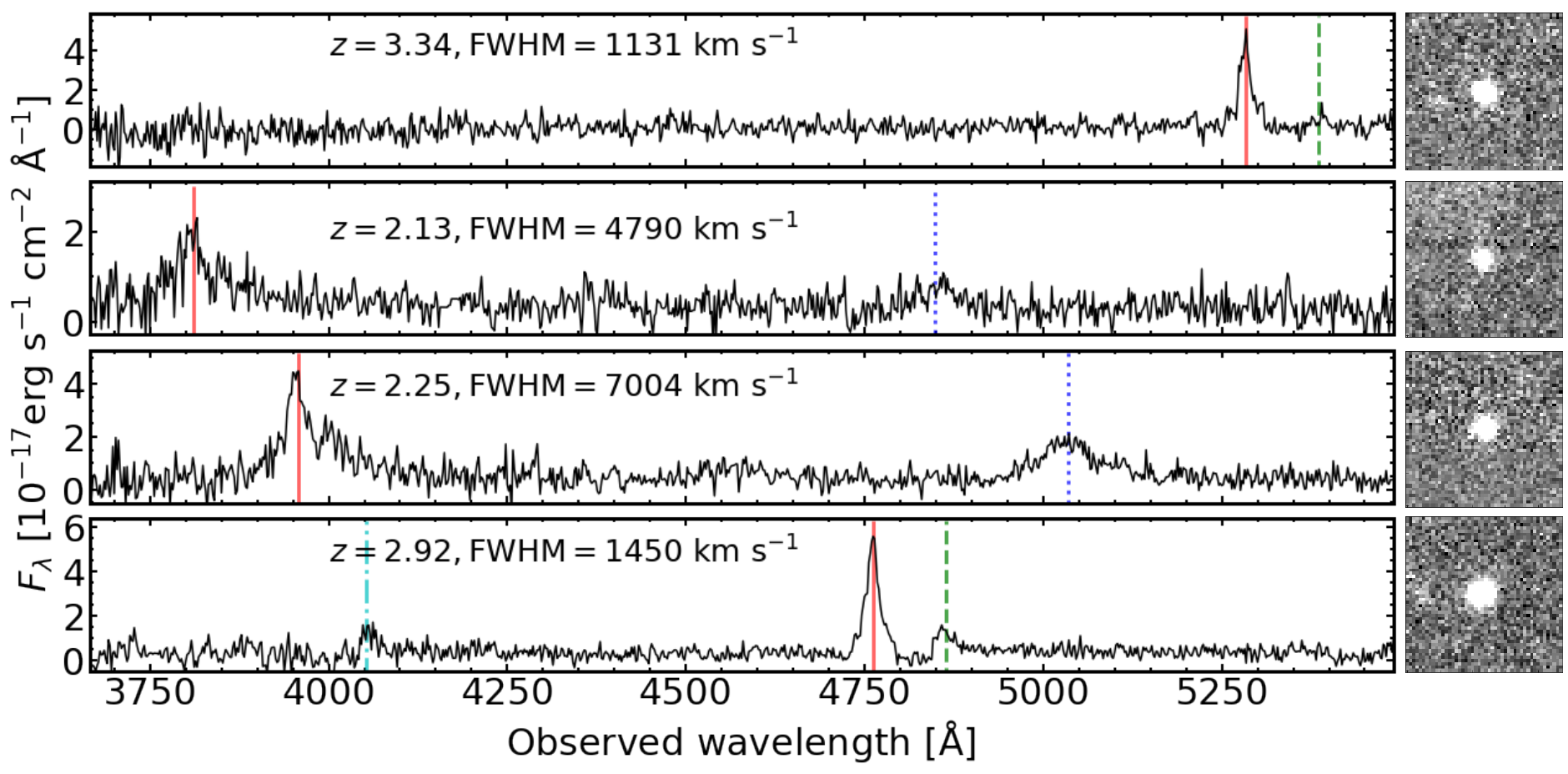}
\end{center}
\caption{Same as Fig.~\ref{fig:nl_laes}, but for our BL-LAEs. Additionally, the blue dotted lines, green dashed lines, and cyan dash-dotted lines represent the wavelengths of detected {\sc Civ}, {\sc Nv} $\lambda 1240$~\AA, and {\sc Ovi} $\lambda 1035$~\AA\ emission lines, respectively. \label{fig:bl_laes}}
\end{figure*}

We examine our BL-LAE selection criteria of $\mathrm{EW_0} > 20$ \AA\ and FWHM $> 1000$ km s$^{-1}$. For type 1 AGN, the major strong permitted and semi-forbidden emission lines within the HETDEX spectral range are Ly$\alpha$, {\sc Civ}, {\sc Ciii}], and Mg{\sc ii}. When a single broad emission line is presented in the HETDEX spectrum, it could be Ly$\alpha$ from AGN at $z=2.0-3.5$,  {\sc Civ} from AGN at $z=1.3-2.0$, {\sc Ciii}] from AGN at $z=0.8-1.3$, or Mg{\sc ii} from AGN at $z=0.3-1.0$. However, our Ly$\alpha$ $\mathrm{EW_0} > 20$ cut would require AGN emitting {\sc Civ}, {\sc Ciii}], and Mg{\sc ii} to have a relatively large rest-frame EWs ($\gtrsim 25$, $31$, and $46$ \AA). Based on stacking analyses of 53 AGN spectra taken with Wide Field Camera 3 on the \textit{HST}, \citet{lusso15} find that the typical rest-frame EWs of {\sc Civ} and {\sc Ciii}] are about $5$ times smaller than that of Ly$\alpha$. The EWs of Mg{\sc ii} are also $\gtrsim 3$ times smaller than those of Ly$\alpha$ for AGN \citep{netzer90}. Thus, the effect of contamination from low-$z$ type 1 AGN would be small. Discussions on the contamination of our BL-LAE subsample is presented in Section \ref{subsec:contam}.
We confirm that $>90\%$ of Ly$\alpha$-emitting AGN at $2.0<z<3.5$ in the SDSS DR14 sample \citep{paris18} satisfy our selection criteria. We obtain the final sample of BL-LAEs that contains $2126$ objects. We notice that $458/2126$ BL-LAEs have secured redshift at $z=2.0-3.5$ based on multiple emission lines identified in HETDEX spectra. Figure \ref{fig:bl_laes} shows four examples of our BL-LAEs.
\begin{figure}[ht!]
\begin{center}
\includegraphics[scale=0.42]{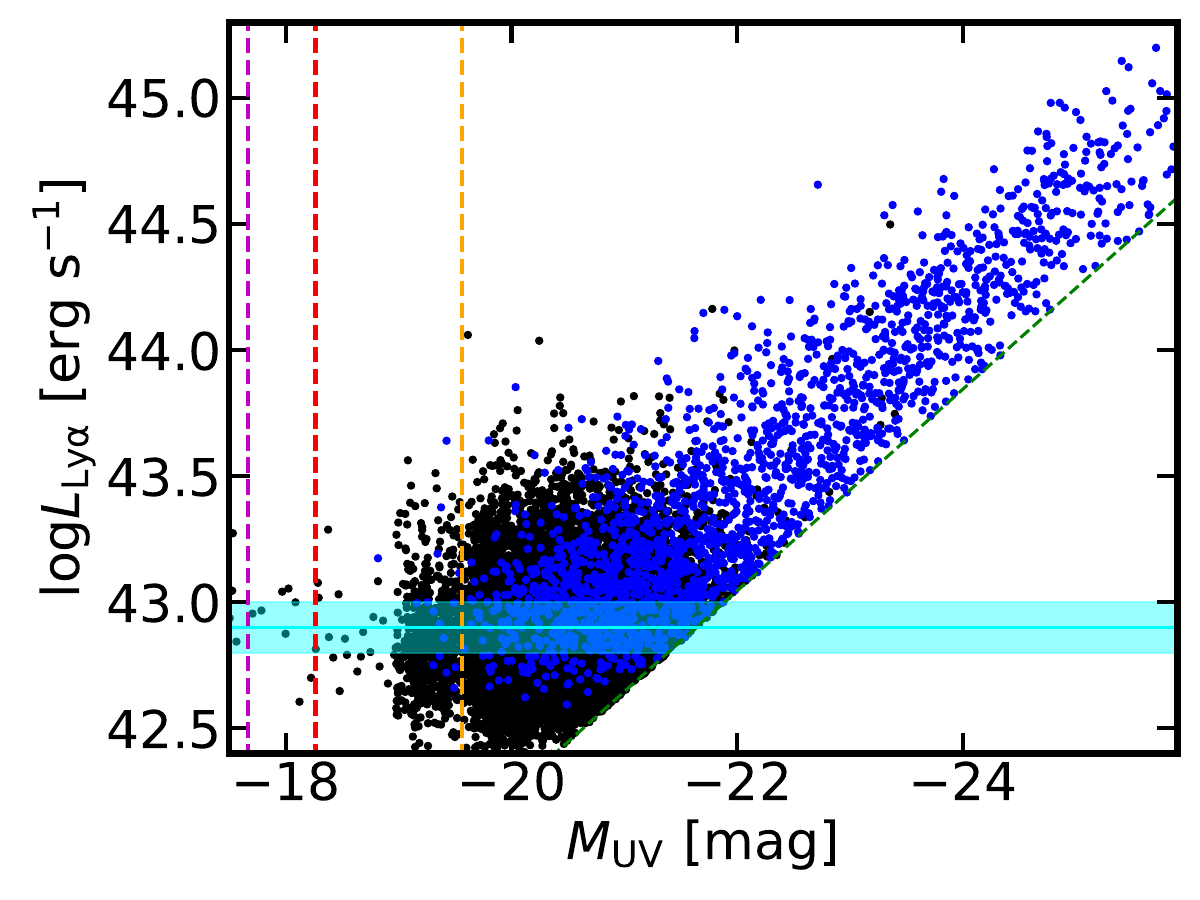}
\end{center}
\caption{$M_\mathrm{UV}-\log L_\mathrm{Ly\alpha}$ distributions of our LAE sample. The blue (black) data points denote BL- (NL-) LAEs. The green dashed line indicates our EW$_0 > 20$ \AA\ cut. The horizontal cyan line corresponds to our $50\%$ $L_\mathrm{Ly\alpha}$ detection limit averaged over our redshift range, with cyan shaded regions indicating the $1\sigma$ uncertainty (68.27\% equal-tailed credible interval). The magenta, red, and orange vertical dashed lines show the $M_\mathrm{UV}$ detection limits in COSMOS, Fall and Spring fields, respectively. \label{fig:Muv_Llya}}
\end{figure}


\subsubsection{Summary of Our LAE Sample} \label{subsec:laesum}
Overall, our LAE selection criteria can be summarized as
\begin{multline}\label{eq:sel_lae}
    \mathrm{EW_0} > 20 \mathrm{\r{A}}\ \mathrm{and} \\ [(\mathrm{FWHM} \leq 1000\ \mathrm{km\ s^{-1}} \ \mathrm{and}\  P_\mathrm{{LAE}}/P_\mathrm{[OII]} > 1) \\ 
    \mathrm{or}\ (\mathrm{FWHM} > 1000\ \mathrm{km\ s^{-1}})]
\end{multline}
With this set of criteria, we select $18320$ LAEs from an effective survey area of $11.4$ deg$^2$. Our LAE sample includes $16194$ ($2126$) NL- (BL-) LAEs (Table \ref{tab:lae}). In Figure \ref{fig:Muv_Llya}, we show the $M_\mathrm{UV}-\log L_\mathrm{Ly\alpha}$ distribution of our LAE sample.

It should be noted that our LAEs defined by Eqation \ref{eq:sel_lae} represent an EW$_0$-limited sample of both Ly$\alpha$-emitting SF galaxies and AGN at the bright $L_\mathrm{Ly\alpha}$ regime ($L_\mathrm{Ly\alpha} \gtrsim 10^{43}$~erg~s$^{-1}$)
that is different from the one of the forthcoming HETDEX studies, which produces a number count offset up to a factor of two.

\subsection{Spectroscopic Follow-up of BL-LAE Candidates}\label{subsec:followup}
To identify the potential {\sc Civ} emission line of our type 1 AGN candidates, we carried out spectroscopic follow-up observations with the DEpt Imaging Multi Object Spectrograph (DEIMOS) on the Keck \Romannum{2} Telescope (PI: Y. Harikane). With the criteria mentioned in Section \ref{subsec:bllae}, we select five BL-LAE candidates that are visible during the observations. Our targets are summarized in Table \ref{tab:deimos_targets}.

The observations were conducted on 2020 February 25 (UT) during the filler time of Harikane et al.'s observations when their main targets were not visible. The seeing size is $0''.4-0''.8$ in FWHM. We used 2 DEIMOS masks, KAKm1 and MUKm1, to cover the 5 objects (Table \ref{tab:deimos_targets}) with the BAL12 filter and the 600ZD grating that is a blue spectroscopy setting for DEIMOS. The slit width was $1''.0$. The spatial pixel scale was $0''.1181\ \mathrm{pixel}^{-1}$. The spectral range is 4500-8000 \AA \ with a resolution of $R \sim 1500$. We took 5 (4) frames for KAKm1 (MUKm1) with a single exposure time of 2,000 seconds. However, one frame for KAKm1 was affected by the high background sky level. We do not use this frame for our analysis. The total number of frames used for our analyses is 4 for each mask, KAKm1 or MUKm1. The effective exposure time is 8,000 seconds. We acquired data of arc lamps and standard star G191B2B. 
\begin{table*}[ht!]
\centering
\caption{Targets of our DEIMOS Observations} \label{tab:deimos_targets}
\begin{tabular}{cccccccc}
\hline
\hline
Object ID & R.A. & Decl. & $z_{\mathrm{Ly\alpha}}$ & $F_{\mathrm{Ly\alpha}}$ & $r$ & $i$ & Note \\
 & (J2000) & (J2000) & & ($10^{-17}$ erg s$^{-1}$ cm$^{-2}$) & (mag) & (mag) & \\
\hline
ID-1 & 14:20:11.8258 & +52:51:50.4864 & 2.33 & 23.04 & 26.4 & 26.5 & No emission line detected \\
ID-2 & 14:19:23.4252 & +52:51:50.4864 & 2.08 & 344.3 & 23.5 & 23.4 & No data available \\
ID-3 & 14:18:46.2857 & +52:41:48.9012 & 2.27 & 92.55 & 23.3 & 23.2 & Foreground galaxy at $z=0.48$ \\
ID-4 & 14:19:28.9898 & +52:49:59.3724 & 2.31 & 67.95 & 22.5 & 22.5 & Type 1 AGN \\
ID-5 & 14:18:33.0739 & +52:43:13.71 & 2.14 & 42.47 & 23.5 & 23.1 & Type 1 AGN \\
\hline
\end{tabular}
\end{table*}

We obtained spectroscopic data listed in Table \ref{tab:deimos_targets}. Because one out of 5 objects, ID-2, unfortunately fall on the broken CCD of DEIMOS, DEIMOS spectra are available for 4 out of 5 objects. Usually, the DEIMOS spectroscopic data are reduced with the \texttt{spec2d} IDL pipeline \citep{davis03}, which performs the bias subtraction, flat fielding, image stacking and wavelength calibration. However, the \texttt{spec2d} pipeline does not work in our blue spectroscopy setting due to the lack of wavelength data in the blue wavelength. We thus carry out the data reduction manually with the \texttt{IRAF} software package \citep{tody86}, using the arc lamps for the wavelength calibration and standard star G91B2B for the flux calibration. We obtain 1D spectra of objects by summing up 15 pixels ($\sim1''.77$) in the spatial direction.

We search for emission lines in both 2D and 1D spectra of the objects listed in \ref{tab:deimos_targets} at the expected wavelengths, using the redshifts derived from the HETDEX Ly$\alpha$ emission lines. We identify {\sc Civ} and {\sc Ciii]} emission lines in two out of the four faint BL-LAE candidates, ID-4 and ID-5. One of the four faint BL-LAE candidates, ID-3, is a foreground galaxy at $z=0.48$ with H$\beta$ and {\sc [Oiii]} emission lines identified in the DEIMOS spectrum. Another BL-LAE candidate, ID-1, has no detectable emission lines in the DEIMOS spectrum. The $3\sigma$ limiting flux of {\sc Civ} ({\sc Ciii]}) of ID-1 is $2.16(1.87)\times 10^{-18}$ erg s$^{-1}$ cm$^{-2}$. Detailed spectroscopic properties of ID4 and ID5 are presented in Section \ref{subsec:bllae_spec}.

\section{Deriving the Luminosity Functions} \label{sec:LF}
\subsection{Contamination} \label{subsec:contam}
We estimate the contamination rate of our NL- ($f_\mathrm{NL}$) and BL-LAEs ($f_\mathrm{BL}$) by cross-matching our sample with spectroscopic and/or photometric catalogs. We also explore the morphology of BL-LAEs in archival \textit{Hubble Space Telescope} (\textit{HST}) imaging data. Because the number of objects available for $f_\mathrm{NL}$ ($f_\mathrm{BL}$) estimation is not sufficient to build a redshift and luminosity correlation, we apply a uniform $f_\mathrm{NL}$ ($f_\mathrm{BL}$) for our NL- (BL-) LAEs. The errors of $f_\mathrm{NL}$ and $f_\mathrm{BL}$ are based on Poisson statistics.

We estimate $f_\mathrm{NL}$ as described in Section \ref{subsec:nllae}. We make a subsample of $32$ NL-LAEs in COSMOS field that are spectroscopically and/or photometrically identified by previous studies. We find that four out of these 32 known galaxies are foreground contaminants. We adopt the $f_\mathrm{NL} = 13\%\pm 8\%$ ($4/32$) for our NL-LAE subsample.

For BL-LAEs, there are two types of contamination. The first type of contamination is the foreground type 1 AGN whose emission lines are redshifted to the HETDEX spectral range (Section \ref{subsec:bllae}). We estimate the fraction of such contamination $f_\mathrm{BL,foreground}$ by crossmatching our BL-LAE subsample with SDSS DR14 QSO catalog \citep{paris18}. Our crossmatching results shows that 111 of our BL-LAEs are previously identified AGN. Among these 111 objects, 99 objects are Ly$\alpha$-emitting AGN at $z=2-3.5$. The remaining 12 objects turn out to be foreground AGN (10 {\sc Civ}-emitting AGN at $z \sim 1.4$, one {\sc Ciii}]-emitting AGN at $z \sim 1.0$, and one Mg{\sc ii}-emitting AGN at $z \sim 0.8$).  We thus apply $f_\mathrm{BL,foreground} = 11\%\pm 3\%$ ($12/111$).

The second type of contamination is the broad emission line mimicked by blended narrow lines emitted from close-galaxy pairs on the sky at the similar wavelengths. We examine the morphology of BL-LAE candidates in COSMOS field with archival high-resolution optical images taken with Advanced Camera for Surveys (ACS) on \textit{Hubble Space Telescope} (\citealt{koekemoer07, massey10a}). We find that three out of 10 BL-LAE candidates have multiple components in the \textit{HST}/ACS images that indicate possible close-galaxy pairs. Because the multiple components do not necessarily all have emission lines, we add errors allowing $f_\mathrm{close}$ to range from $0$ to $30\%+20\%$ (3/10). $f_\mathrm{BL}$ is obtained by:
\begin{equation} \label{eq:lum}
    f_\mathrm{BL} = f_\mathrm{foreground} + f_\mathrm{close}.
\end{equation}

\subsection{Detection Completeness} \label{subsec:comp}
\begin{figure}[ht!]
\begin{center}
\includegraphics[scale=0.37]{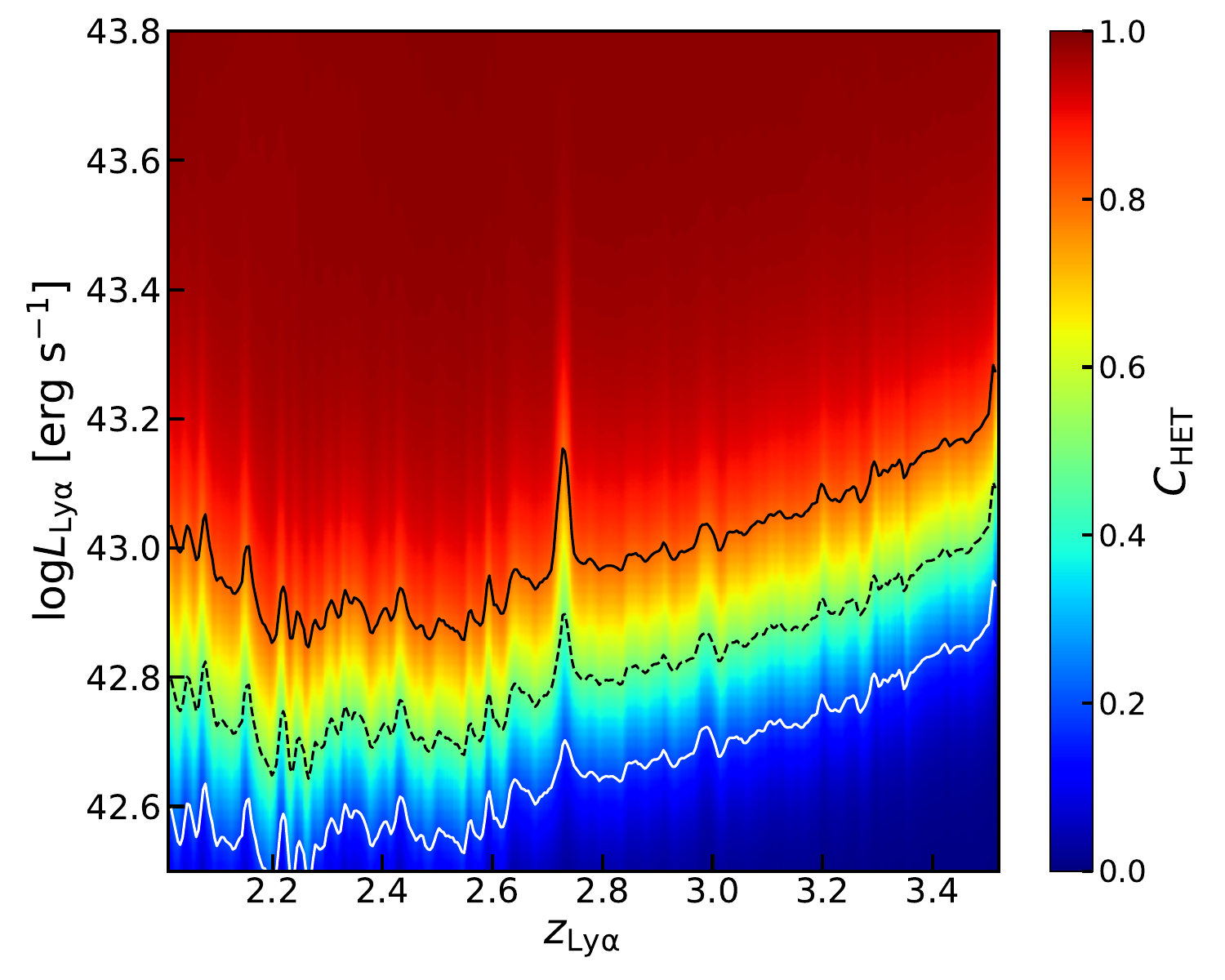}
\end{center}
\caption{Completeness map of our HETDEX LAEs. The solid white line, dashed black line, and solid black line indicate the $20\%$, $50\%$, and $80\%$ completeness levels, respectively. \label{fig:comp_nl}}
\end{figure}

For LAEs selected from HETDEX emission line catalog, the detection completeness ($C_\mathrm{HET}$) as a function of observed wavelength ($\lambda$) and emission line flux ($F$) is estimated from simulated LAEs inserted into real HETDEX data (Farrow et al., in preparation; Gebhardt et al., in preparation). From $\lambda$ and $F$, we calculate Ly$\alpha$ redshift $z$ and luminosity $L_\mathrm{Ly\alpha}$,
transforming $C_\mathrm{HET}(\lambda,F)$ to $C_\mathrm{HET}(z,L_\mathrm{Ly\alpha})$. Figure \ref{fig:comp_nl} presents an example of $C_\mathrm{HET}(z,L_\mathrm{Ly\alpha})$.

For HSC-detected LAEs, the detection completeness $C_\mathrm{HSC}$ consists of two completeness values of the HSC $r$-band source detection $C_\mathrm{HSC,cont}$ and the HETDEX emission line detection $C_\mathrm{HSC,line}$, where
\begin{equation} \label{eq:comp_bl}
    C_\mathrm{HSC}(z, M_\mathrm{UV}, L_\mathrm{Ly\alpha}) = C_\mathrm{HSC,cont}(z, M_\mathrm{UV}) \times C_\mathrm{HSC,line}(z, L_\mathrm{Ly\alpha}).
\end{equation}

\begin{figure}[ht!]
\begin{center}
\includegraphics[scale=0.4]{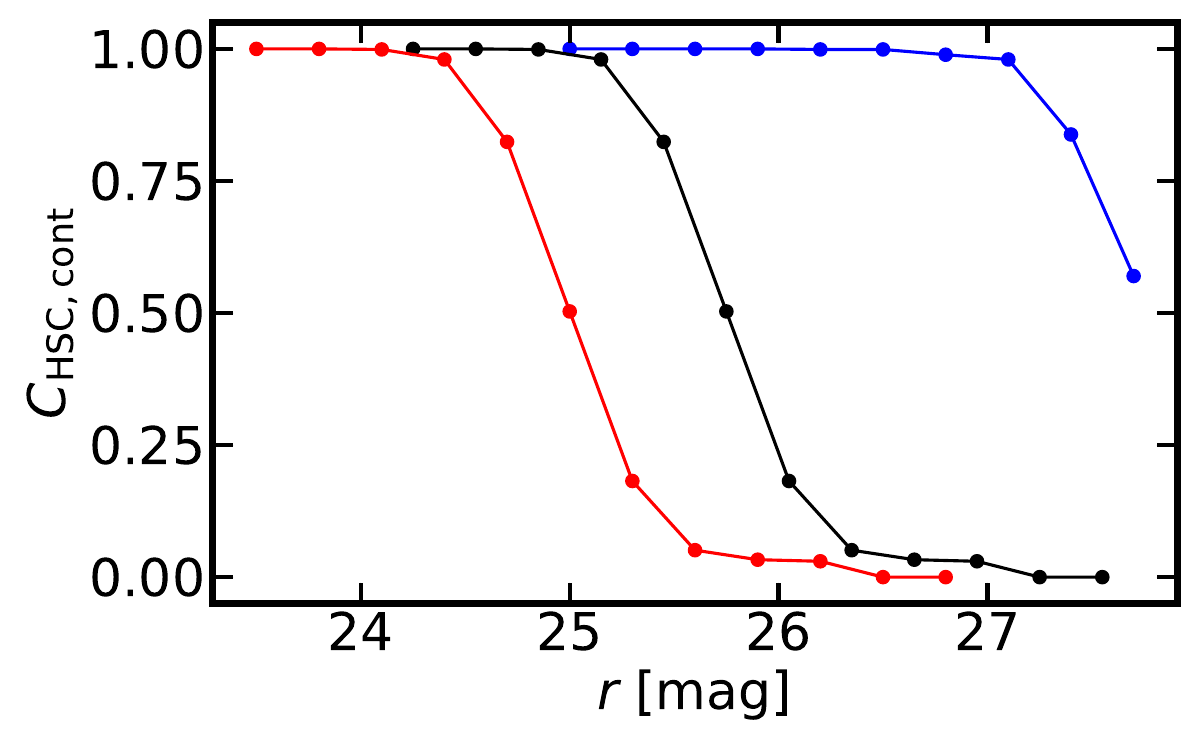}
\end{center}
\caption{Completeness of our HSC $r$-band imaging data. The blue, black, and red line denote the completeness of imaging data in the COSMOS, Fall, and Spring field, respectively. \label{fig:comp_hsc}}
\end{figure}

We derive $C_\mathrm{HSC,cont}$ from \citet{kakuma20}, who has calculated the completeness of their HSC-SSP $r$-band imaging data as a function of magnitude. We scale their completeness function by the difference of the $5\sigma$ limiting magnitudes between their HSC SSP wide field data ($r_{5\sigma}=25.8$) and our HETDEX-HSC data ($r_{5\sigma}=25.1$), estimating the completeness function of $C_\mathrm{HSC_cont}$ that is shown in Figure \ref{fig:comp_hsc}. 

\begin{figure}[ht!]
\begin{center}
\includegraphics[scale=0.37]{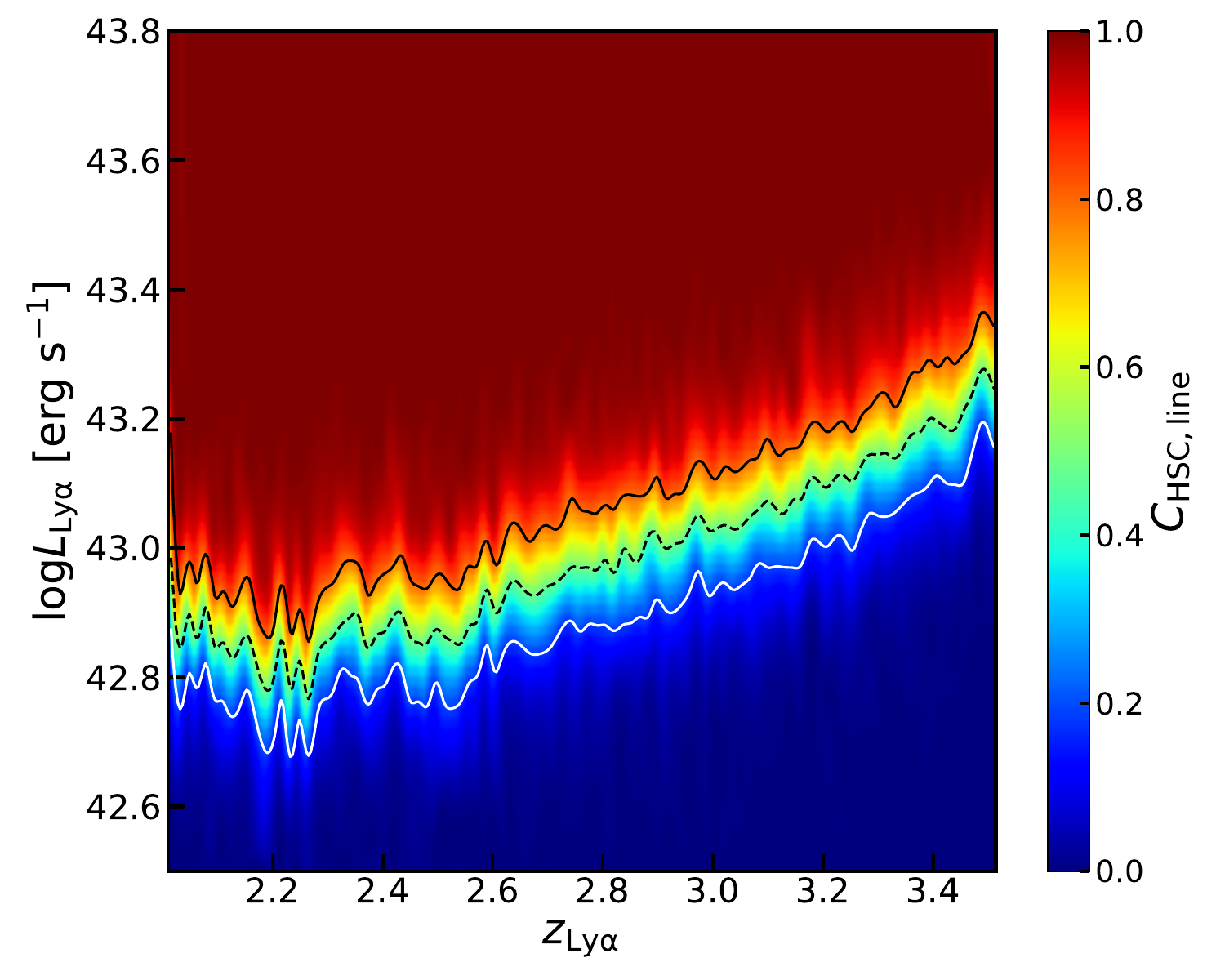}
\end{center}
\caption{Same as Figure \ref{fig:comp_nl}, but for our our HSC-detected LAEs.\label{fig:comp_bl}}
\end{figure}

We estimate $C_\mathrm{HSC,line}$ by Monte Carlo simulations. Specifically, we make $500$ mock spectra with Gaussian line models. For each object in our sample, we fix the FWHM of the line models to the emission line fitting result given in Section \ref{subsec:hsc_laes}. We then produce composite fiber spectra, assuming the 2D light distribution is described by the Moffat PSF with a $2''$ FWHM value. We add noise to the composite fiber spectra, randomly generating the noise following a Gaussian distribution function whose sigma is the median value of the noise in the fiber spectra. We perform the emission-line detection on the mock spectra in the same manner as Section \ref{subsec:hsc_laes}, and calculate $C_\mathrm{HSC,line}$ that is defined by the fraction of the detected artificial lines to the total mock spectra. We repeat this process with various $z$ and $L_\mathrm{Ly\alpha}$, obtaining $C_\mathrm{HSC,line}(z, L_\mathrm{Lya})$. An example of $C_\mathrm{HSC,line}$ is shown in Figure \ref{fig:comp_bl}.

\subsection{Ly$\alpha$ Luminosity Function} \label{subsec:LF_lya}
\begin{figure*}[ht!]
\begin{center}
\includegraphics[scale=0.6]{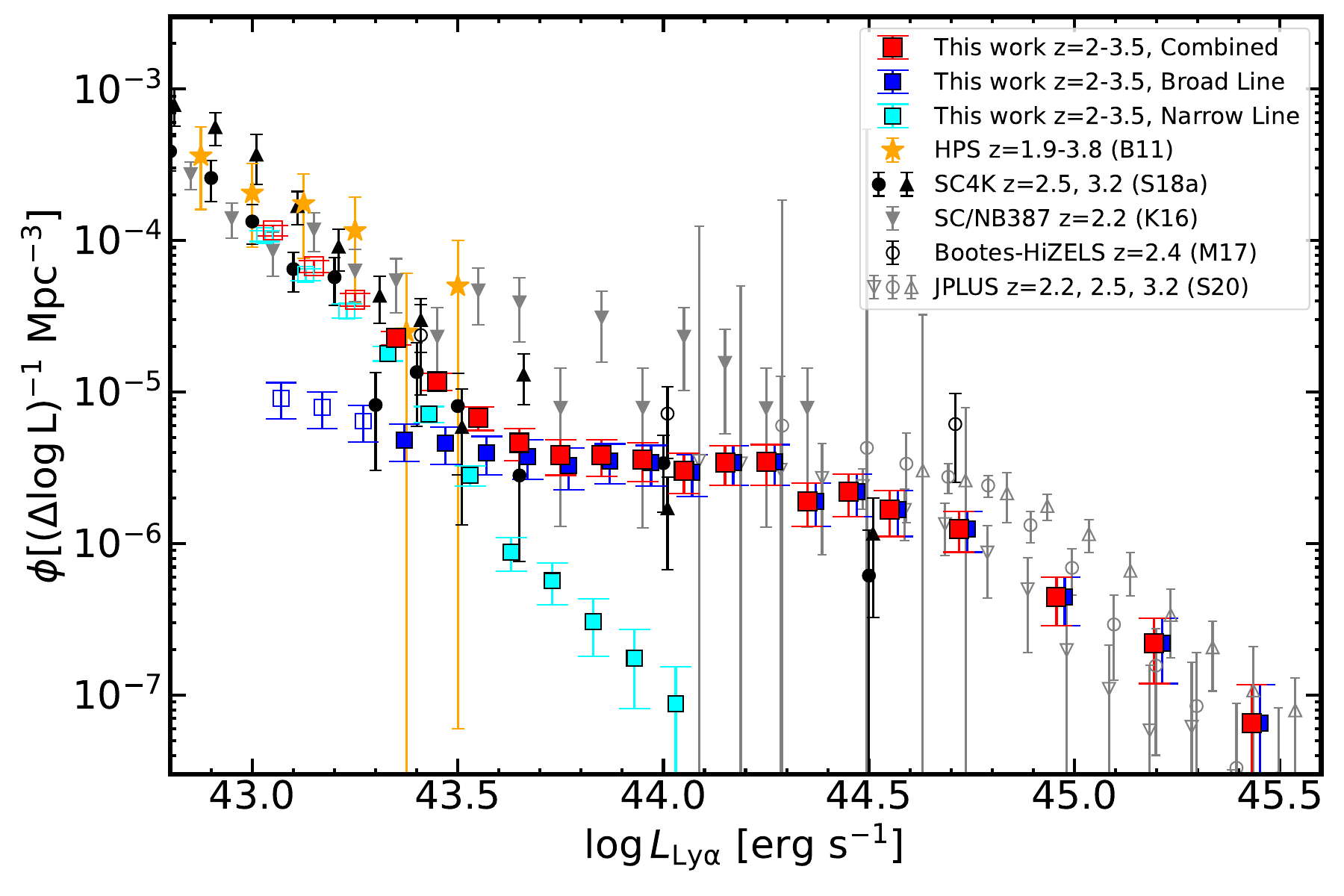}
\end{center}
\caption{Ly$\alpha$ LF of our NL- (cyan squares), BL- (blue squares) and C-LAEs (red squares) within the $11.4$~deg$^2$ survey area. Data points that may affected by incompleteness due to Eddington bias is marked with open squares (see text). For clarity, we slightly shift the data points of NL- and BL-LAEs along the abscissa. The orange stars indicate the results from previous spectroscopic study of \citet{blanc11} at $1.9 < z < 3.8$ (B11). We also show the Ly$\alpha$ LFs at $\log L_\mathrm{Ly\alpha}[\mathrm{erg~s^{-1}}] \gtrsim 43.3$ derived by previous photometric surveys/studies at three redshift slices of $z$ = 2.2 (downwards triangles), 2.4-2.5 (circles), and 3.2 (upwards triangles) that are taken from \citet{konno16} (K16, grey solid symbols), \citet{matthee17} (M17, black open symbols), \citet{sobral18a} (S18a, black solid symbols), and \citet{spinoso20} (S20, grey open symbols). \label{fig:LF}}
\end{figure*}
We derive the Ly$\alpha$ LF with our LAE sample. We apply the non-parametric $1/V_\mathrm{max}$ estimator \citep{schmidt68,felten76} that accounts for the redshift and luminosity dependent selection function of our LAE sample. For the $i$-th object in our LAE sample,  $V_{\mathrm{max},i}$ corresponds to the maximum comoving volume inside which it is detectable. The $V_{\mathrm{max},i}$ value depends on the detection completeness functions $C_{\mathrm{HET}}(z,L_{\mathrm{Ly\alpha}})$ and $C_{\mathrm{HSC}}(z,M_{\mathrm{UV}},L_{\mathrm{Ly\alpha}})$,
\begin{equation} \label{eq:vmax}
    V_{\mathrm{max},i}=\omega \int_{z_\mathrm{min}}^{z_\mathrm{max}}C_{i} \frac{\mathrm{d}V}{\mathrm{d}z}\mathrm{d}z,
\end{equation}
where $\omega$ and $\frac{dV}{dz}$ denote the angular area of the survey and the differential comoving volume element, respectively. Here, $z_\mathrm{min}$ ($z_\mathrm{max}$) is the lower (upper) limit of the redshift range of the survey. We calculate $V_\mathrm{max}$ of each NL- (BL-) LAE, $V^\mathrm{NL}_\mathrm{max}$ ($V^\mathrm{BL}_\mathrm{max}$), and obtain the number densities of NL- and BL-LAEs in each luminosity bin with:
\begin{equation} \label{eq:lf_vmaxnl}
    \phi_{\mathrm{NL}}(\log L_\mathrm{Ly\alpha})=\frac{1-f_{\mathrm{NL}}}{\Delta(\log L_\mathrm{Ly\alpha})}\sum_k \frac{1}{V^{\mathrm{NL}}_{\mathrm{max},k}} ,
\end{equation}
\begin{equation} \label{eq:lf_vmaxbl}
    \phi_{\mathrm{BL}}(\log L_\mathrm{Ly\alpha})=\frac{1-f_{\mathrm{BL}}}{\Delta(\log L_\mathrm{Ly\alpha})}\sum_k \frac{1}{V^{\mathrm{BL}}_{\mathrm{max},k}} ,
\end{equation}
where $\Delta(\log L_\mathrm{Ly\alpha})$ is the luminosity bin width and $k$ is the number of objects in each luminosity bin. The summations in Eqs. (\ref{eq:lf_vmaxnl}) and (\ref{eq:lf_vmaxbl}) are performed over the $k$ objects in each luminosity bin. 

Figure \ref{fig:LF} shows the binned Ly$\alpha$ LF of our LAE sample at $43.0 < \log L_\mathrm{Ly\alpha}[\mathrm{erg~s^{-1}}] < \mathbf{45.5}$, where our LF reaches $50\%$ completeness. Note that at $\log L_\mathrm{Ly\alpha} < 43.3$~erg~s$^{-1}$, our Ly$\alpha$ LFs may suffer from a non-negligible Eddington bias \citep{eddington13} that is generated from uncertainties in HETDEX spectro-photometry (Gebhardt et al., in preparation). We hence do not include the Ly$\alpha$ LF data points at $43.0 < \log L_\mathrm{Ly\alpha}/[\mathrm{erg~s^{-1}}] < 43.3$ into our analysis. The error bars of our results are estimated from the quadrature sum of the Poissonian errors, errors in contamination correction, and flux measurement errors. \footnote{We also evaluate the errors in luminosity distance measurements by considering the Ly$\alpha$ velocity offsets ($\Delta v_\mathrm{Ly\alpha}$). Typical Ly$\alpha$-emitting galaxies have $\Delta v_\mathrm{Ly\alpha}\sim 200-300$~km~s$^{-1}$ (e.g., \citealt{verhamme18}). For AGNs at $z\sim 2-3$, we investigate $\Delta v_\mathrm{Ly\alpha}$ based on SDSS DR14 QSO catalog \citep{rakshit20}. We find the 16$^{th}$, 50$^{th}$, and 84$^{th}$ percentiles of the $\Delta v_\mathrm{Ly\alpha}$ distribution to be -301, 75, and 598~km~s$^{-1}$, respectively. Assuming a $\Delta v_\mathrm{Ly\alpha}=598$~km~s$^{-1}$ would result in the uncertainty of $<1\%$ in the luminosity, which is negligible compared with the bin widths of our LFs. We hence do not consider the errors in luminosity distances.} Our LFs span a wide Ly$\alpha$ luminosity range of $43.3 < \log L_\mathrm{Ly\alpha}/[\mathrm{erg~s^{-1}}] < \mathbf{45.5}$, showing a significant bright-end excess at $\log L_\mathrm{Ly\alpha}/[\mathrm{erg~s^{-1}}]\gtrsim 43.5$ that is dominated by BL-LAEs who are type 1 AGN with FWHM(Ly$\alpha$)$>$1000~km~s$^{-1}$. 

We compare our results in Figure \ref{fig:LF} with those from previous studies at the similar redshift and $L_\mathrm{Ly\alpha}$ ranges. At $z\sim 2-3$, \citet{blanc11} is the only work that presents the spectroscopically-derived Ly$\alpha$ LF at $\log L_\mathrm{Ly\alpha}/[\mathrm{erg~s^{-1}}] \gtrsim 43$. Due to the limited survey area of 169~arcmin$^2$ and the small sample of 80 LAEs, their LF does not show a bright-end hump and has large errors in the two brightest $L_\mathrm{Ly\alpha}$ bins at $\log L_\mathrm{Ly\alpha}/[\mathrm{erg~s^{-1}}] \sim 43.3-43.6$. Over this $L_\mathrm{Ly\alpha}$ range, our results are consistent with theirs while having significantly smaller errors. At $\log L_\mathrm{Ly\alpha}/[\mathrm{erg~s^{-1}}] \sim 43.3-44$, our Ly$\alpha$ LFs are comparable with previous photometric studies (e.g., \citealt{konno16,matthee17,sobral18a}) given the relatively large scatter in their results, which can possibly be attributed to the different small redshift intervals probed by these NB surveys. At the brightest end with $\log L_\mathrm{Ly\alpha}/[\mathrm{erg~s^{-1}}] \gtrsim 44$ where BL-LAEs dominate, our LF aligns well with \citet{spinoso20}, supporting their suggestion that the bright LAEs in their sample are QSOs.
\begin{figure*}[ht!]
\begin{center}
\includegraphics[scale=0.47]{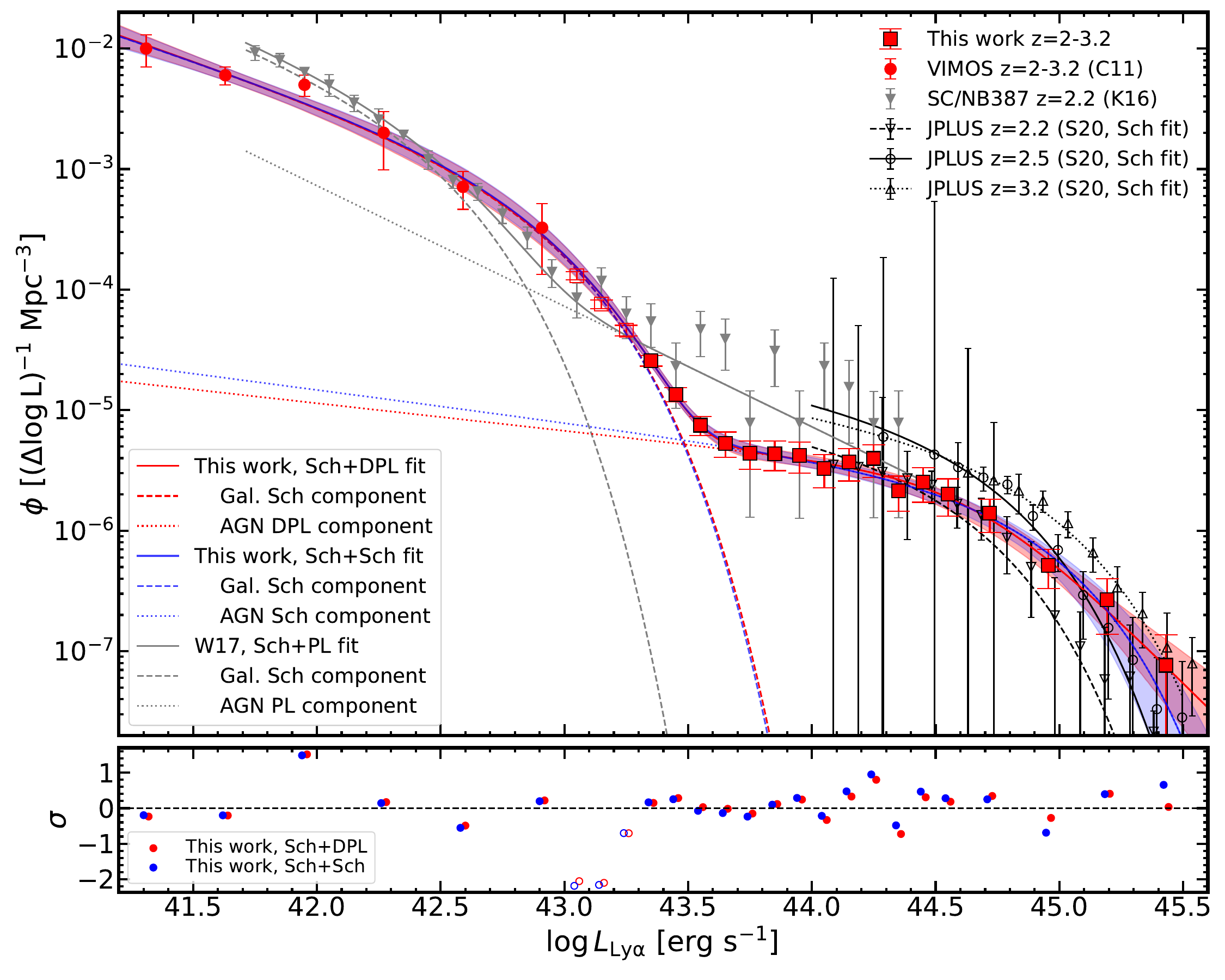}
\end{center}
\caption{Top: Best-fit Ly$\alpha$ LF of our LAE sample with $2 < z < 3.2$ (red squares) and \citet{cassata11} at $2.0 < z < 3.2$ (red circles). The red (blue) line shows our best-fit result of Model 1(2) with the measurements of our study and \citet{cassata11}, with shaded region corresponding to the $1\sigma$ uncertainty (68.27\% equal-tailed credible interval). The black open downward triangles (black dashed line), clrcles (solid line), and upward triangles (dotted line) represents the binned Ly$\alpha$ LFs (best Schechter fit) from \citet{spinoso20} at $z \sim$ 2.25, 2.54, and 3.24, respectively. The grey triangles indicate the results from \citet{konno16}. We also show in grey lines the fitting results from \citet{wold17} on the obseved Ly$\alpha$ LF of \citet{konno16}. The dashed, dotted, and solid grey lines indicate their best-fit Schechter component, power law component, and the overall LF, respectively. Bottom: The residuals of Model 1 (red) and 2 (blue) fit in units of the uncertainty in each luminosity bin. \label{fig:LF_fit}} 
\end{figure*}

We conduct model fitting with our Ly$\alpha$ LF in Figure \ref{fig:LF}. In general, Ly$\alpha$ LFs at $\log L_\mathrm{Ly\alpha} \lesssim 43.5$~erg~s$^{-1}$ are well described by the Schechter function. Because our Ly$\alpha$ LF does not cover $\log L_\mathrm{Ly\alpha}[\mathrm{erg~s^{-1}}] < 43.3$, we constrain the faint end of the Ly$\alpha$ LF by including the binned Ly$\alpha$ LF at $z=2-3.2$ from \citet{cassata11} in our fitting. The Ly$\alpha$ LF of \citet{cassata11} is derived with spectroscopically identified LAEs whose $L_\mathrm{Ly\alpha}$ range from 10$^{41}-10^{43}$~erg~s$^{-1}$, hence complementary to our sample. 

To parameterize the shape of the bright-end hump in our Ly$\alpha$ LF, we considering the following two models separately. In Model 1, we assume the bright-end follows a double power law (DPL) that is broadly used to describe the QSO/AGN UV LF (e.g., \citealt{boyle88,pei95,boyle00,richards06a,stevans18}). The DPL is defined as
\begin{multline} \label{eq:pl}
    \phi_\mathrm{DPL}(L_\mathrm{Ly\alpha}) \mathrm{dlog}L_\mathrm{Ly\alpha} =  \ln10\ \phi^*_\mathrm{DPL} \\
    \left(\frac{L_\mathrm{Ly\alpha}}{L^*_\mathrm{DPL}}\right) \left[\left(\frac{L_\mathrm{Ly\alpha}}{L^*_\mathrm{DPL}}\right)^{-\alpha_\mathrm{DPL}}+\left(\frac{L_\mathrm{Ly\alpha}}{L^*_\mathrm{DPL}}\right)^{-\beta_\mathrm{DPL}}\right]^{-1}\mathrm{dlog}L_\mathrm{Ly\alpha},
\end{multline}
where $L^*_\mathrm{DPL}$, $\phi^*_\mathrm{DPL}$, $\alpha_\mathrm{DPL}$, and $\beta_\mathrm{DPL}$ are characteristic luminosity, the normalization factor, faint-end slope, and bright-end slope, respectively. In Model 2, we assume the bright-end can be described by a Schechter function as suggested by \citet{spinoso20}.
\begin{table*}[ht!]
\begin{center}
\caption{Fitting results of our Ly$\alpha$ LF}
\label{tab:LF_fit}
\centering
\begin{tabular}{cccccccccc}
\hline
\hline
\multirow{3}{*}{Model} &
\multicolumn{3}{c}{Faint Component} & &
\multicolumn{4}{c}{Bright Component} \\\cline{2-4}\cline{6-9}
& $\alpha$ & $\log L^*$ & $\log \phi^*$ & & $\alpha$ & $\beta$ & $\log L^*$ & $\log \phi^*$ \\
& & [erg s$^{-1}$] & [Mpc$^{-3}$] & & & & [erg s$^{-1}$] & [Mpc$^{-3}$] \\
\hline
Model 1 (Schechter$+$DPL) & $-1.70_{-0.14}^{+0.13}$ & $42.87_{-0.07}^{+0.09}$ & $-3.41_{-0.24}^{+0.20}$ & & $-1.23_{-0.24}^{+0.42}$ & $-3.06_{-0.90}^{+0.57}$ &
$44.68_{-0.40}^{+0.26}$ & $-5.92_{-0.32}^{+0.26}$ \\
\hline
Model 2 (Schechter$+$Schechter) & $-1.69_{-0.14}^{+0.13}$ & $42.86_{-0.07}^{+0.09}$ & $-3.39_{-0.23}^{+0.19}$ & & $-1.27_{-0.19}^{+0.23}$ & -- &
$44.85_{-0.19}^{+0.19}$ & $-5.96_{-0.27}^{+0.24}$ \\
\hline
\end{tabular}
\end{center}
\end{table*}

We fit our binned Ly$\alpha$ LF with the function of
\begin{equation} \label{eq:schdpl}
    \phi(L_\mathrm{Ly\alpha}) = \phi_\mathrm{Faint}(L_\mathrm{Ly\alpha}) + \phi_\mathrm{Bright}(L_\mathrm{Ly\alpha}),
\end{equation}
where $\phi_\mathrm{Faint}$ represents the Schechter function for the faint component, and $\phi_\mathrm{Bright}$ is the DPL (Schechter) function of Model 1(2) for the bright component.
We obtain posterior probability distributions for the free parameters in Equation \ref{eq:schdpl} using the Markov Chain Monte Carlo (MCMC) technique. For the prior of $L^*_\mathrm{DPL}$ and $L^*_\mathrm{Sch}$ in $\phi_\mathrm{Bright}$, we assume uniform distributions that is larger than $L^*_\mathrm{Sch}$ in $\phi_\mathrm{Faint}$ and smaller than the maximum observed $L_\mathrm{Ly\alpha}$ ($10^\mathbf{45.5}$~erg~s$^{-1}$) in our Ly$\alpha$ LF. Since the number of our BL-LAEs at the luminosity brighter than the break of DPL in Model 1 is not large, we adopt a uniform prior for $\beta_\mathrm{DPL}$ in Model 1 covering a reletively narrow range of $[-5,-2]$ (e.g.,\citealt{kul19}). We assume broad, uniform priors for the other parameters. We use the {\tt emcee} code \citep{fm13} for MCMC. For each parameter, we adopt the posterior median as our best-fit value and the 68.27\% equal-tailed credible interval as the uncertainty. The fitting results of our two models and the best-fit functions are presented in Table \ref{tab:LF_fit} and Figures \ref{fig:LF_fit}, respectively. The posterior probability distributions of the parameters of Model 1(2) are shown in Figure \ref{fig:LF_fit_contour} (\ref{fig:LF_fit_contour_m2}). We find that Model 1 can better describe the decay in our Ly$\alpha$ LF at $\log L_\mathrm{Ly\alpha}[\mathrm{erg~s^{-1}}] \gtrsim 44.8$ with smaller residuals, although the differences are smaller than 1$\sigma$. To compare the performance of the two models, We also calculate the difference in Bayesian information criterion \citep{schwarz78} that is defined as
\begin{equation} \label{eq:bic}
    \Delta\mathrm{BIC} = \chi^2_2 - \chi^2_1 + (k_2 - k_1)\ln N.
\end{equation}
In Equation \ref{eq:bic}, $\chi^2_1$($\chi^2_2$) and $k_1$($k_2$) is the $\chi^2$ and the number of free parameters of Model 1(2), respectively, and $N$ represents the number of data points that are used in the fitting. We find a $\Delta\mathrm{BIC}$ value of -2.17, suggesting that Model 1 is slightly favored over Model 2. We hence choose Model 1 (Schechter + DPL fit) as the best-fit model of our Ly$\alpha$ LF in the following discussions, although we cannot rule out a Schechter exponential decay of the Ly$\alpha$ LF at $\log L_\mathrm{Ly\alpha}[\mathrm{erg~s^{-1}}] \gtrsim 45.8$ as described by Model 2.

We compare our best-fit model of Ly$\alpha$ LFs with previous studies. For the bright DPL component, our faint-end slope $\alpha_\mathrm{DPL}$ is consistent with the results from \citet{spinoso20}, who found an weighted-average $\alpha$ value of $-1.35\pm 0.84$ over their four redshift slices.  For the faint component, our results have reasonably well constrained Shechter parameters $L^*_\mathrm{Sch}$, $\alpha_\mathrm{Sch}$, and $\phi^*_\mathrm{Sch}$ that are consistent with \citet{cassata11}. We note that \citet{cassata11} fit their LF with a fixed $L^*$ due to the lack of data at the bright end ($L_\mathrm{Ly\alpha} > 10^{43}$~erg~s$^{-1}$). Our results provide strong constraints at $L_\mathrm{Ly\alpha} > L^*_\mathrm{Sch}$, determining the best-fit Shechter function by fitting three Schechter parameters simultaneously.

\begin{figure}[h!]
\begin{center}
\includegraphics[scale=0.4]{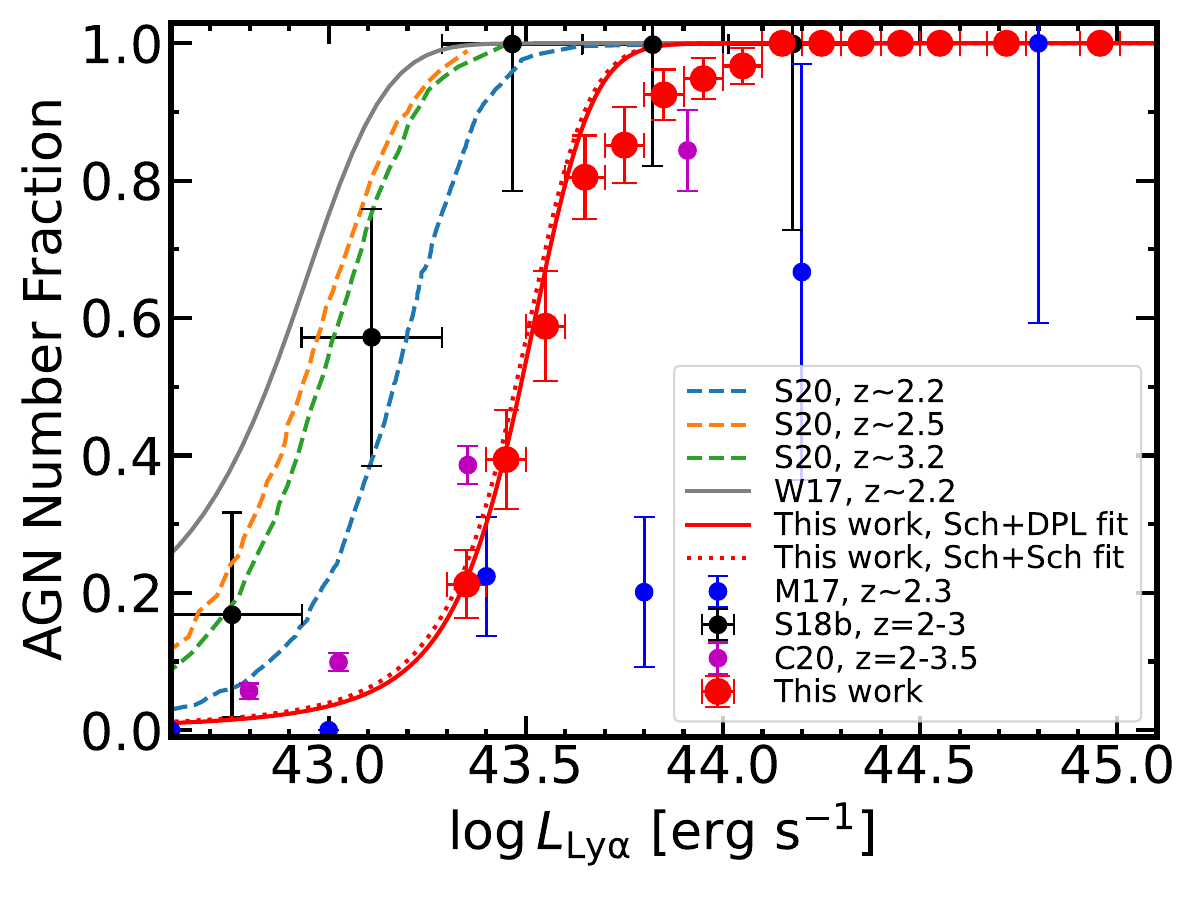}
\end{center}
\caption{AGN number fraction as a function of $\log L_\mathrm{Ly\alpha}$. Our results derived with Equation \ref{eq:agnfrac_lya} are shown in the red circles. The functional form of our $f_{\mathrm{AGN,Ly\alpha}}$ derived with Equation \ref{eq:agnfrac_lyafit} using Model 1(2) is presented in the red solid (dashed) line. The other symbols represent previous measurements whose references are listed in the legend. M17: \citet{matthee17}; W17: \citet{wold17}; S18: \citet{sobral18b}; C20: \citet{calhau20}.  \label{fig:fagn_lya}}
\end{figure}

From our binned and best-fit Ly$\alpha$ LFs, we investigate the type 1 AGN number fraction ($f_{\mathrm{AGN,Ly\alpha}}$) of our LAE sample. Note that our $f_{\mathrm{AGN,Ly\alpha}}$ only accounts for type 1 AGN and hence represents the lower limit. First, we calculate $f_{\mathrm{AGN,Ly\alpha}}$ with
\begin{equation} \label{eq:agnfrac_lya}
    f_{\mathrm{AGN,Ly\alpha}}(\log L_\mathrm{Ly\alpha})=\frac{\phi_{\mathrm{BL}}(\log L_\mathrm{Ly\alpha})}{\phi_{\mathrm{BL}}(\log L_\mathrm{Ly\alpha})+\phi_{\mathrm{NL}}(\log L_\mathrm{Ly\alpha})},
\end{equation}
where $\phi_\mathrm{NL}$ and $\phi_\mathrm{BL}$ are the binned LFs defined in Equation \ref{eq:lf_vmaxnl} and \ref{eq:lf_vmaxbl}, respectively. We present our results with those estimated with various types of AGN in Figure \ref{fig:fagn_lya}. Our $f_{\mathrm{AGN,Ly\alpha}}$ increases rapidly with $L_\mathrm{Ly\alpha}$, rising from $\sim 20\%$ to $\sim 90-100\%$ from $\log L_\mathrm{Ly\alpha}/[\mathrm{erg~s^{-1}}] \sim 43.3$ to $\sim 43.8$. Such a trend is in good agreement with the results derived with radio and X-ray detected AGN \citep{calhau20}. \citet{matthee17} also found a similar increase in $f_{\mathrm{AGN,Ly\alpha}}(L_\mathrm{Ly\alpha})$ using X-ray detected LAEs. At $\log L_\mathrm{Ly\alpha}/[\mathrm{erg~s^{-1}}] \sim 43.6-44.4$, their results are lower than ours, which may be attributed to their relatively shallow X-ray and Ly$\alpha$ data \citep{calhau20}. Another study that discuss such an increase in $f_{\mathrm{AGN,Ly\alpha}}(L_\mathrm{Ly\alpha})$ is \citet{sobral18b} based on 21 spectroscopically identified objects that include both type 1 and type 2 AGN. At $\log L_\mathrm{Ly\alpha} \leq\ 43.5$~erg~s$^{-1}$, our $f_{\mathrm{AGN,Ly\alpha}}$ are lower than their results. The small sample and inclusion of type 2 AGN may explain the larger $f_{\mathrm{AGN,Ly\alpha}}$ from \citet{sobral18b}. 

Second, we estimate $f_{\mathrm{AGN,Ly\alpha}}$ with our best-fit Ly$\alpha$ LF, assuming that the faint Schechter component (bright DPL component) can well describe the Ly$\alpha$ LFs of SF galaxies (AGN). The similar method have been applied by \citet{wold17} and \citet{spinoso20} to derive $f_{\mathrm{AGN,Ly\alpha}}(L_\mathrm{Ly\alpha})$. However, as mentioned in \citet{spinoso20}, $f_{\mathrm{AGN,Ly\alpha}}(L_\mathrm{Ly\alpha})$ estimated with this method only represents an illustrative results, given the strong assumption and the high sensitivity to the determination of the faint Schechter component. We calculate $f_{\mathrm{AGN,Ly\alpha}}$ with 
\begin{equation} \label{eq:agnfrac_lyafit}
    f_{\mathrm{AGN,Ly\alpha}}(\log L_\mathrm{Ly\alpha})=\frac{\phi_{\mathrm{Bright}}(\log L_\mathrm{Ly\alpha})}{\phi_{\mathrm{Bright}}(\log L_\mathrm{Ly\alpha})+\phi_{\mathrm{Faint}}(\log L_\mathrm{Ly\alpha})}.
\end{equation}
As shown in Figure \ref{fig:fagn_lya}, our $f_{\mathrm{AGN,Ly\alpha}}(L_\mathrm{Ly\alpha})$ estimated with Equation \ref{eq:agnfrac_lyafit} agrees nicely with the ones derived with Equation \ref{eq:agnfrac_lya}. Comparing with the results from \citet{spinoso20} that are based on their bright AGN/QSO LFs and the faint SF galaxy LFs from \citet{sobral18a}, our $f_{\mathrm{AGN,Ly\alpha}}(L_\mathrm{Ly\alpha})$ has a similar increase, and shifts towards brighter $L_\mathrm{Ly\alpha}$ by $\Delta\log L_\mathrm{Ly\alpha}/[\mathrm{erg~s^{-1}}] \sim 0.3-0.5$. \citet{wold17} obtained a $f_{\mathrm{AGN,Ly\alpha}}(L_\mathrm{Ly\alpha})$ similar to \citet{spinoso20}, but with a flatter increase, by fitting the observed LF from \citet{konno16} that has large scatters at the bright end. They also estimated the observed Ly$\alpha$ luminosity density contributed from SF galaxies ($\rho^\mathrm{obs}_\mathrm{Ly\alpha,SF}$) and AGN ($\rho^\mathrm{obs}_\mathrm{Ly\alpha,AGN}$) to be $10^{39.7}$ and $10^{39.5}$~erg~s$^{-1}$~Mpc$^{-3}$, respectively, by integrating their best-fit Ly$\alpha$ LFs over the Ly$\alpha$ luminosity range of $\log L_\mathrm{Ly\alpha} = 41.41-44.4$~erg~s$^{-1}$. They concluded that at $z=2.2$, AGN contribute to $\sim 40\%$ to the total observed Ly$\alpha$ luminosity density, which is consistent with their observational results at $z=0.3$ and $z=0.9$. We follow their method, calculating $\rho^\mathrm{obs}_\mathrm{Ly\alpha,SF}$ and $\rho^\mathrm{obs}_\mathrm{Ly\alpha,AGN}$ from our best-fit Ly$\alpha$ LF with
\begin{equation}\label{eq:rhosf}
    \rho^\mathrm{obs}_\mathrm{Ly\alpha,SF} = \int L_\mathrm{Ly\alpha}\phi_\mathrm{Bright}(L_\mathrm{Ly\alpha})\mathrm{d}L_\mathrm{Ly\alpha},
\end{equation}
\begin{equation}\label{eq:rhoagn}
    \rho^\mathrm{obs}_\mathrm{Ly\alpha,AGN} = \int L_\mathrm{Ly\alpha}\phi_\mathrm{Faint}(L_\mathrm{Ly\alpha})\mathrm{d}L_\mathrm{Ly\alpha}.
\end{equation}
We integrate Equations (\ref{eq:rhosf}) and (\ref{eq:rhoagn}) over the same $L_\mathrm{Ly\alpha}$ range as mentioned in \citet{wold17}, obtaining the $\rho^\mathrm{obs}_\mathrm{Ly\alpha,SF}$ ($\rho^\mathrm{obs}_\mathrm{Ly\alpha,AGN}$) value of $10^{39.7}$ ($10^{38.6}$)~erg~s$^{-1}$~Mpc$^{-3}$. Despite the different shape of best-fit Schechter function as presented in Figure \ref{fig:LF_fit}, our $\rho^\mathrm{obs}_\mathrm{Ly\alpha,SF}$ is consistent with the result from \citet{wold17}. Contrastingly, our $\rho^\mathrm{obs}_\mathrm{Ly\alpha,AGN}$ is $\sim 10$ times lower, resulting in an AGN contribution of $\sim 8\%$ to the total Ly$\alpha$ luminosity density at $z=2.0-3.2$.


\subsection{UV LF}
\label{subsec:LF_uv}
\begin{figure*}[ht!]
\begin{center}
\includegraphics[scale=0.5]{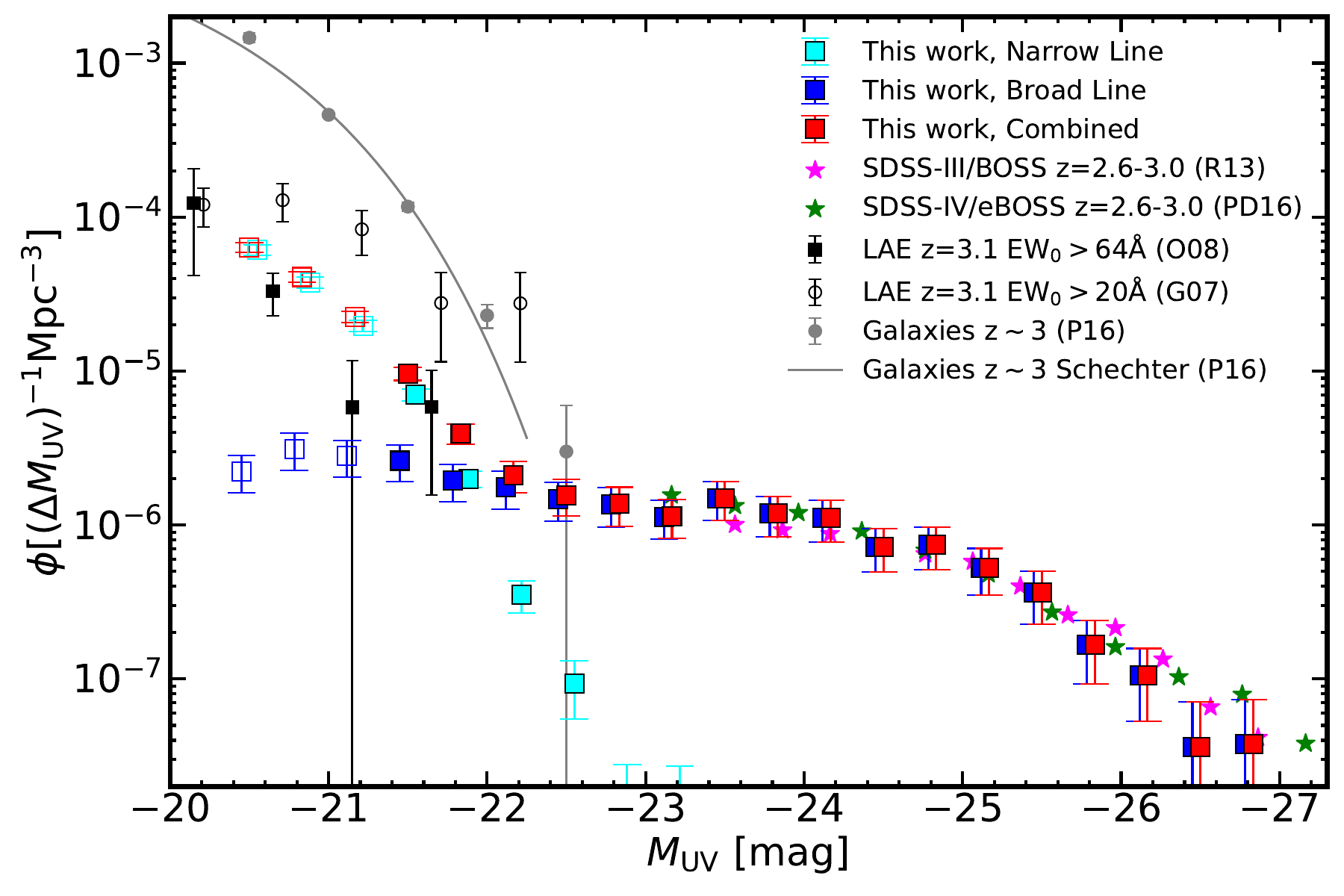}
\end{center}
\caption{UV LF of our NL- (cyan squares), BL- (blue squares), and C-LAEs (red squares). Data points that may affected by incompleteness is marked with open squares. The magenta and green stars denote the QSO UV LF at $z\sim 2.8$ from SDSS-III (\citealt{ross13}; R13) and SDSS-IV (\citealt{pd16}; PD16), respectively. The black squares and black open circles indicate the LAE UV LFs from \citet[O08]{ouchi08} and \citet[G07]{gronwall07}, respectively. We also show the UV LF of photometrically selected galaxies at $z\sim 3$ in the grey circles and solid line (\citealt{parsa16}; P16). \label{fig:LFuv}}
\end{figure*}

We derive the UV LF of our LAE sample in the same manner as described in Section \ref{subsec:LF_lya}. We convert the $r$-band magnitude ($r$) of our LAEs to the absolute UV magnitude ($M_\mathrm{UV}$) using the following relation:
\begin{equation} \label{eq:r_muv}
    M_\mathrm{UV} = r - 5\log( \frac{d_\mathrm{L}}{10\mathrm{pc}}) + 2.5\log(1+z) + K,
\end{equation}
where $d_\mathrm{L}$ and $z$ are the luminosity distance in pc and the redshift of our LAEs, respectively, determined by the Ly$\alpha$ emission lines in HETDEX spectra. We set the $K$-correction term ($K$) to be 0 according to our assumption of flat UV continua (Section\ref{subsec:laesel}). Figure \ref{fig:LFuv} presents our UV LF at $-27 < M_\mathrm{UV} < -21$, where the completeness reaches $50\%$. At the bright end with $M_\mathrm{UV} < -22.5$, our results agree well with the SDSS QSO UV LF \citep{ross13,pd16}. Similar to our Ly$\alpha$ LF, the bright end of our UV LF are dominated by BL-LAEs, indicating the presence of type 1 AGN. At the faint end ($M_\mathrm{UV} > -21.5$), our UV LF enters the SF galaxy regime where previous results on the UV LFs of LAEs and photometrically selected are available. Our results are generally consistent with previous UV LFs of LAEs derived by \citet{ouchi08} and \citet{gronwall07}. Our LAE sample may suffer incompleteness at $M_\mathrm{UV} > -21.3$ due to the detection limit of Ly$\alpha$ emission lines (Figure \ref{fig:Muv_Llya}). As a result, only LAEs with strongest Ly$\alpha$ emission lines are included in our sample. We denote our UV LF at $M_\mathrm{UV} > -21.3$ with open symbols. 
\par
\begin{figure}[ht!]
\begin{center}
\includegraphics[scale=0.4]{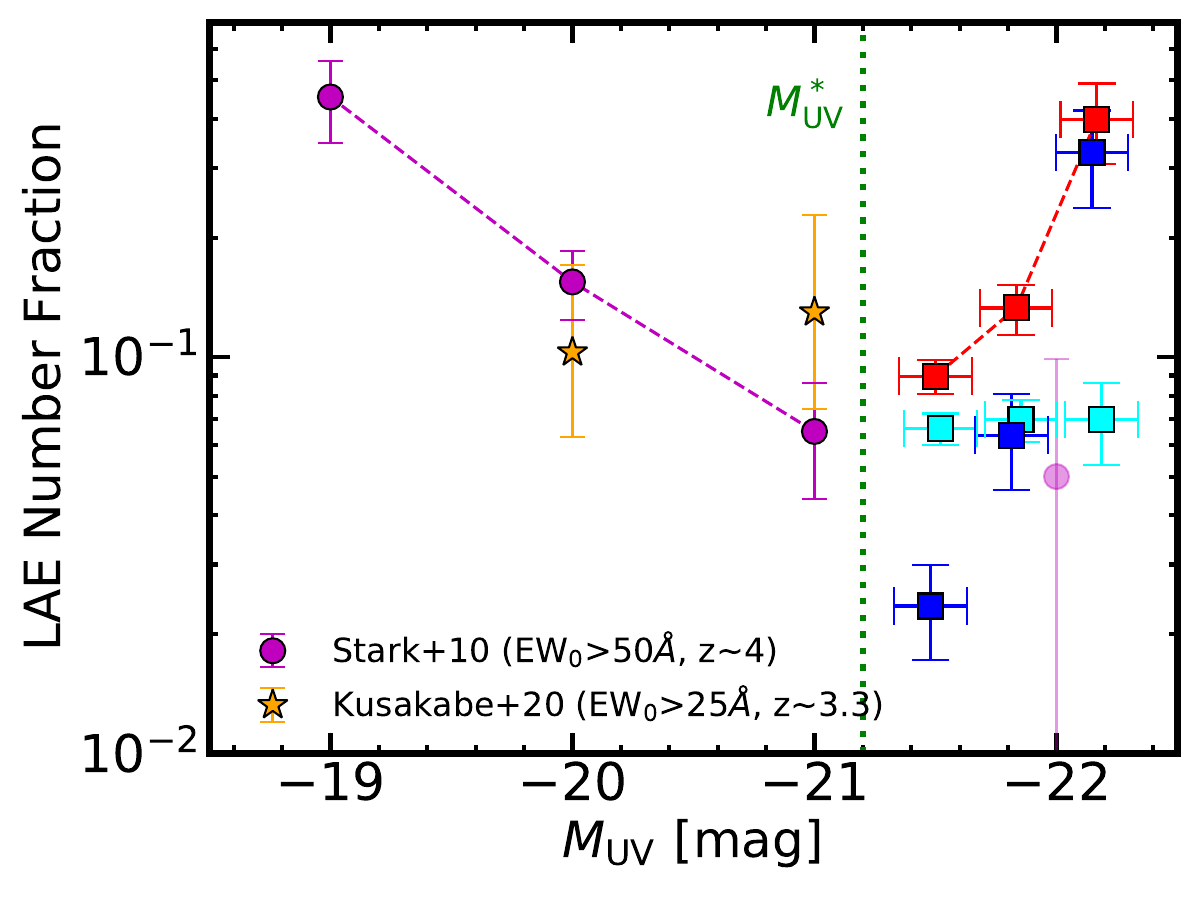}
\end{center}
\caption{$X_\mathrm{LAE}$ as a function of $M_\mathrm{UV}$. The red, blue, and cyan squares denote our results derived from C-, BL-, and NL-LAEs, respectively. The magenta circles and yellow stars represent the results from \citet{stark10} and \citet{kusakabe20}, respectively. At $M_\mathrm{UV}$ fainter (brighter) than $M^*_\mathrm{UV} \sim -21$, $X_\mathrm{LAE}$ increases (decreases) with $M_\mathrm{UV}$. \label{fig:fagn_uv}}
\end{figure}

We study the LAE number fraction ($X_\mathrm{LAE}$) as a function of $M_\mathrm{UV}$ based on our LAE UV LF ($\phi_\mathrm{UV,LAE}$) and the best-fit UV LF of galaxies at $z~\sim~3$ ($\phi_\mathrm{UV,z=3}$) derived by \citet{parsa16}. They obtained their galaxy sample based on multiband photometry with the photometric redshift determined by SED fitting. By assuming the galaxy sample of \citet{parsa16} is complete, we calculate $X_\mathrm{LAE}$ with:
\begin{equation} \label{eq:x_lae}
    X_\mathrm{LAE}(M_\mathrm{UV}) = \frac{\phi_\mathrm{UV,LAE}(M_\mathrm{UV})}{\phi_\mathrm{UV,z=3}(M_\mathrm{UV})}.
\end{equation}
In Figure \ref{fig:fagn_uv}, we present our results covering $-22.1<M_\mathrm{UV}<-21.2$, where both our complete LAE UV LF and the galaxy UV LF are available. At $M_\mathrm{UV}<-21$, our results show that $X_\mathrm{LAE}$ increases towards brighter UV luminosity. Such a trend is mainly due to BL-LAEs, whereas the number fraction of NL-LAEs does not change at $-22.5\lesssim M_\mathrm{UV}\lesssim -21$. We compare our results with the $X_\mathrm{LAE}$ values derived by two previous studies based on LBG samples at the similar redshift, \citet{stark10} and \citet{kusakabe20}, both of which focused on SF galaxies without AGN. 
\citet{stark10} showed that $X_\mathrm{LAE}$ decreases with UV luminosity at $-22.5<M_\mathrm{UV}<-18.5$. Their $X_\mathrm{LAE}$ at $M_\mathrm{UV}=-22$ is in agreement with our results derived from NL-LAEs alone. Contrastingly, \citet{kusakabe20} found no UV luminosity dependence of $X_\mathrm{LAE}$ at $M_\mathrm{UV}\sim -20$ and $-21$, though their values are consistent with those in \citet{stark10} within errors. Such a discrepancy may be caused by the increasing LBG selection bias towards faint $M_\mathrm{UV}$ \citep{kusakabe20}. Our results suggest that while the number fraction of SF-dominated LAEs has a tendency of decrease towards brighter UV luminosity, the number fraction of Ly$\alpha$-emitting type 1 AGN becomes dominant at $M_\mathrm{UV}\lesssim -21$ and increases towards brighter UV luminosity. This results in a turnaround of $X_\mathrm{LAE}$ at $M_\mathrm{UV}\sim -21$.  
It should be noted that the UV LF of BL-LAEs shown here is calculated from the total UV flux that includes both AGN and stellar components of host galaxies. In Section \ref{subsec:uvLF_evolution}, we evaluate the contributions of UV-continuum emission from AGN and their host galaxies based on SED fitting results from \citet{kakuma20}, and discuss the redshift evolution of AGN activities.

\section{Spectroscopic Properties of BL-LAEs}
\label{subsec:bllae_spec}
\begin{figure*}[ht!]
\begin{center}
\includegraphics[scale=0.55]{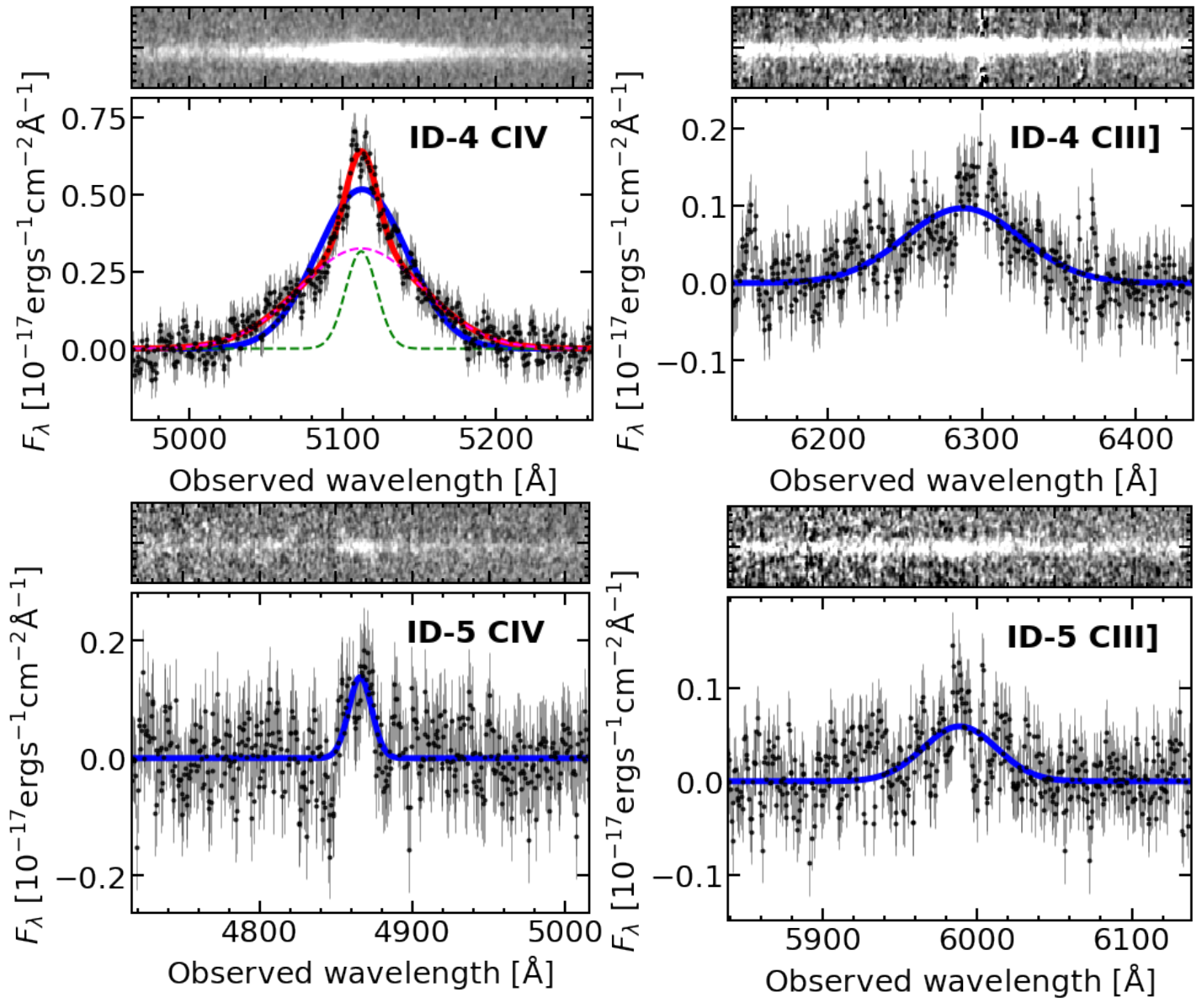}
\end{center}
\caption{Emission line spectra of ID-4 and ID-5. At the top of each panel, we show the 2D spectrum. At the bottom of each panel, we present the 1D spectrum (black circles with errors) and the best-fit Gaussian profile (blue line). In the top-right panel, we also show the best-fit double Gaussian profile (red line) as well as the broad (cyan dotted line) and the narrow (green dotted line) component. \label{fig:linefit}}
\end{figure*}

In this Section, we present spectroscopic properties of the two faint BL-LAEs with spectra, ID-4 and ID-5 (Table \ref{tab:deimos_targets}). We measure the central wavelengths, fluxes, and FWHMs of the {\sc Civ} and {\sc Ciii]} emission lines (Figure \ref{fig:linefit}) in the spectra of ID-4 and ID-5 with a best-fit Gaussian profile. We estimate the errors of fluxes and FWHMs by the Monte Carlo simulations. For each emission line, we make 1000 mock emission line spectra by adding noise to the observed line spectrum. The noise added to the observed line spectrum is generated, following a Gaussian probability distribution whose standard deviation is the $1\sigma$ uncertainty in the observed line spectrum. The errors of the measured fluxes and FWHMs are defined as the $68\%$ confidence intervals in the distributions of the fluxes and FWHMs in the mock emission line spectra. We measure the FWHMs of {\sc Civ} and {\sc Ciii]}, FWHM$_\mathrm{CIV}$ and FWHM$_\mathrm{CIII]}$, to be $4043.2_{-136.4}^{+138.8}$ $(1072.2_{-104.4}^{+96.8})$ and $4268.0_{-402.4}^{+373.7}$ $(2024.1_{-1202.6}^{+485.6})$ km s$^{-1}$, respectively, for ID-4 (ID-5). For the {\sc Civ} emission line of ID-4, we find a broad component that cannot be well fitted with a single Gaussian profile. We fit the {\sc Civ} emission line of ID-4 with a double Gaussian profile that consists of a broad and a narrow components, measuring the FWHM with the broad component of the double Gaussian profile. We obtain FWHM$_\mathrm{CIV}$ of the broad component is $5536.3_{-185.8}^{+202.1}$ km s$^{-1}$. 
Table \ref{tab:em_properties} summarizes the emission line properties of our objects.

\begin{table*}[ht!]
\begin{threeparttable}[t]
\centering
\caption{Emission line properties} \label{tab:em_properties}
\begin{tabular}{cccccc} 
\hline
\hline
Object & Emission Line & Wavelength & Flux & FWHM & Reduced $\chi^2$ \\
& & (\AA) & ($10^{-17}$ erg s$^{-1}$ cm$^{-2}$) & (km s$^{-1}$) & \\
\hline
\hline
\hline
\multirow{2}{4em}{ID-4} & {\sc Civ} & $5112.9$ & $37.83_{-1.00}^{+1.06}$ & $4043.2_{-136.4}^{+138.8}$ ($5536.3_{-185.8}^{+202.1}$)\tnote{a} & $1.157$ ($0.861$)\tnote{b} \\ 
& {\sc Ciii]} & $6287.7$ & $9.19_{-0.75}^{+0.77}$ & $4268.0_{-402.4}^{+373.7}$ & $0.943$ \\ 
\hline
\multirow{2}{4em}{ID-5} & {\sc Civ} & $4865.8$ & $2.50_{-0.34}^{+0.35}$ & $1072.2_{-104.4}^{+95.8}$ & $1.043$ \\ 
& {\sc Ciii]} & $5988.6$ & $2.50_{-1.00}^{+0.57}$ & $2024.1_{-1202.6}^{+485.6}$ & $1.326$ \\ 
\hline
\hline
\end{tabular}
\begin{tablenotes}
\item[a] The number in the parenthesis indicate the FWHMs of the broad component in the two-component Gaussian fit.
\item[b] The number in the parenthesis indicate the reduced $\chi^2$ of the two-component Gaussian fit.
 \end{tablenotes}
\end{threeparttable}
\end{table*}

We use the scaling relations of \citet{kim18} with $\mathrm{FWHM_{{\sc Civ}}}$ and the monochromatic continuum luminosity at rest-frame 1350\AA\ ($L_\mathrm{1350}$) to derive the black hole masses $M_\mathrm{BH}$ of ID-4 and ID-5. We estimate $L_\mathrm{1350}$ by the log-linear extrapolation from the observed $r$-band and $i$-band fluxes as listed in Table \ref{tab:deimos_targets}. The $M_\mathrm{BH}$ value is obtained with:
\begin{equation} \label{eq:Mhb}
    \log\left(\frac{M_\mathrm{BH}}{M_\odot}\right) = A + \log\left\{\left(\frac{L_\mathrm{1350}}{10^{44}\ \mathrm{erg\ s}^{-1}}\right)^\beta\left(\frac{\mathrm{FWHM_{CIV}}}{1000\ \mathrm{km\ s}^{-1}}\right)^\gamma \right\}.
\end{equation}
We adopt the parameters $(A,\beta,\gamma)$ of $(6.7,0.5,2.0)$ that are taken from \citet{kim18}. Our two faint BL-LAEs, ID-4 and ID-5, have the black hole masses of $\log (M_\mathrm{BH}/M_\odot) = 8.70_{-0.03}^{+0.03}$ and $6.85_{-0.08}^{+0.08}$, respectively. 

\begin{figure}[h!]
\begin{center}
\includegraphics[scale=0.4]{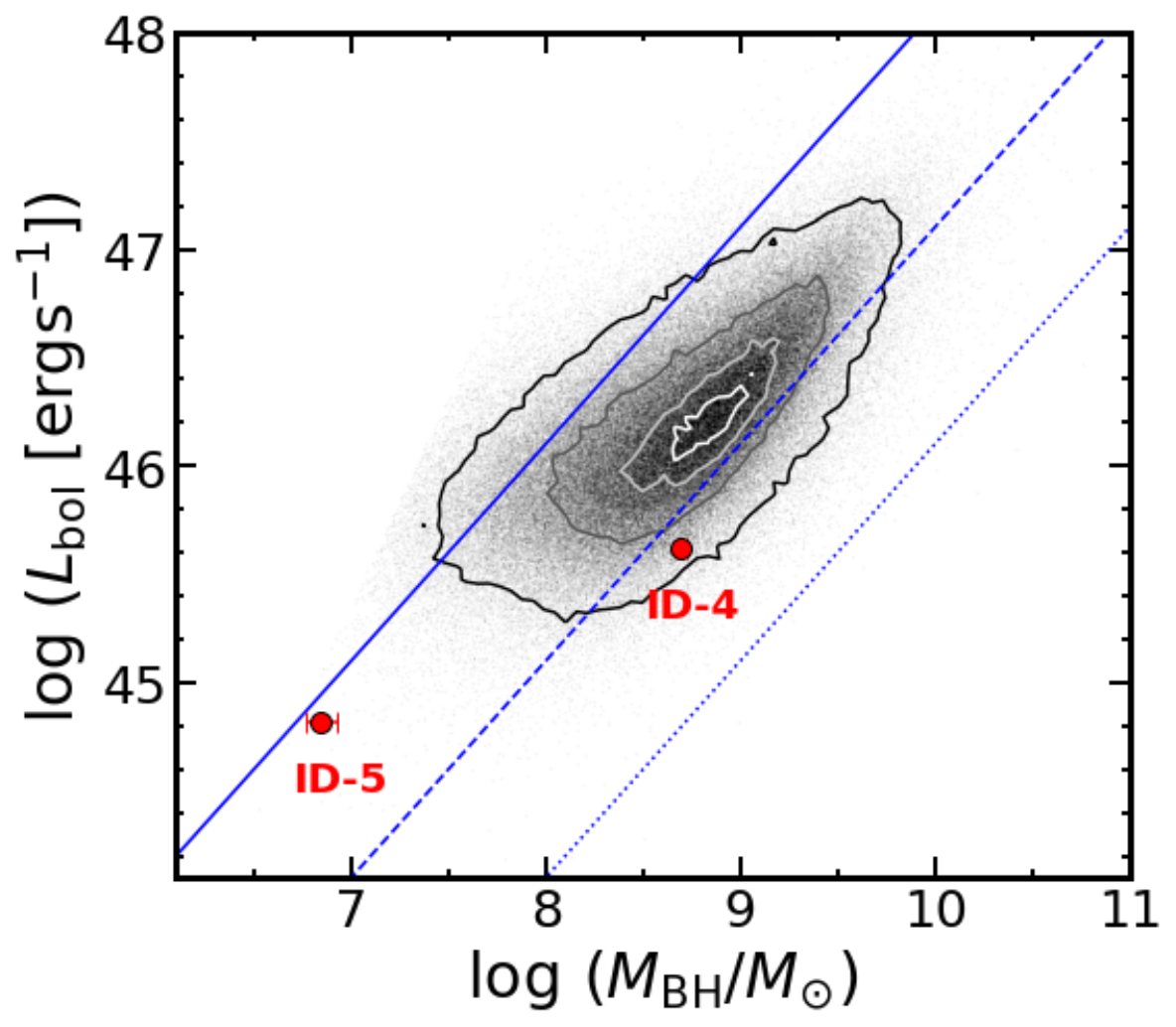}
\end{center}
\caption{$M_\mathrm{BH}-L_\mathrm{bol}$ relation for ID-4 and ID-5. The red circles show our measurements. The grey data points and contours represent the results of $z=2-3.5$ quasars in SDSS DR14 \citep{rakshit20}. The blue solid, dashed and dotted line correspond to the Eddington ratio $\lambda_\mathrm{Edd} = 1, 0.1$ and $0.01$, respectively.  \label{fig:edd_ratio}}
\end{figure}

From the $M_\mathrm{BH}$ values, we calculate the Eddington ratio ($\lambda_\mathrm{Edd}$) of a black hole that is defined as:
\begin{equation} \label{eq:Edd}
    \lambda_\mathrm{Edd} = L_\mathrm{bol} / L_\mathrm{Edd},
\end{equation}
where $L_\mathrm{bol}$ ($L_\mathrm{Edd}$) is the bolometric (Eddington) luminosity of the black hole. The $L_\mathrm{bol}$ are derived from $L_{1350} $ with the bolometric correction factor presented in \citet{richards06} and \citet{shen11}:
\begin{equation} \label{eq:Lbol}
    L_\mathrm{bol} = 3.81 \times L_\mathrm{1350}.
\end{equation}
We derive the $L_\mathrm{bol}$ values of $10^{45.62}$ ($10^{44.82}$) erg s$^{-1}$ for ID-4 (ID-5). The $L_\mathrm{Edd}$ is defined as:
\begin{equation} \label{eq:Ledd}
    L_\mathrm{Edd} = 1.26 \times 10^{38} \left(\frac{M_\mathrm{BH}}{M_\odot}\right).
\end{equation}
The Eddington ratios of ID-4 and ID-5 are $\lambda_\mathrm{Edd} = 0.067_{-0.004}^{+0.005}$ and $=0.746_{-0.121}^{+0.111}$, respectively. Figure \ref{fig:edd_ratio} presents the $M_\mathrm{BH}-L_\mathrm{bol}$ relation for ID-4 and ID-5. 

\section{Discussion}
\subsection{Faint end slope of Ly$\alpha$ LF}\label{subsec:az}

\begin{figure}[ht!]
\begin{center}
\includegraphics[scale=0.4]{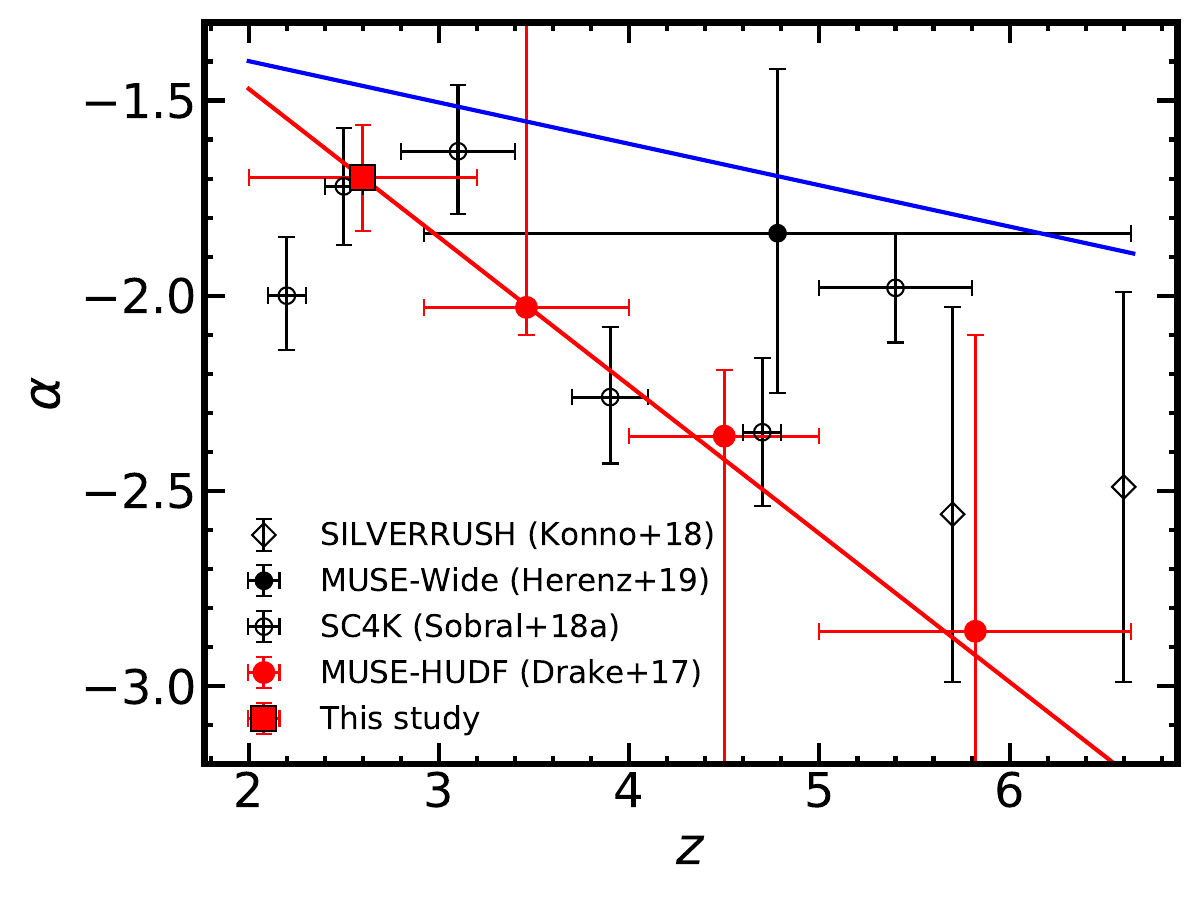}
\end{center}
\caption{Evolution of the faint end slope of Ly$\alpha$ LF. Our result is denoted by the red square. The red circles, black filled circles, black open circles, and black open diamonds indicate the results from \citet{drake17b}, \citet{herenz19}, \citet{sobral18a}, and \citet{konno18}, respectively. The error bars in the abscissa are the redshift ranges of the measurements. Our best-fit linear function is presented in the red line. The blue solid line shows the redshift evolution of the faint end slope of UV LF derived with photometrically selected galaxies \citep{parsa16}.  \label{fig:a_z}}
\end{figure}

Although \citet{gronke15} have predicted the evolution of the faint end slope $\alpha_\mathrm{Sch}$ of the Ly$\alpha$ LF based on a phenomenological model, such evolution has yet to be confirmed by observations. Combining our Ly$\alpha$ LF at $L_\mathrm{Ly\alpha} \geq 10^{43.3}$~erg~s$^{-1}$ and the one derived by \citet{cassata11} at $L_\mathrm{Ly\alpha} \leq 10^{43}$~erg~s$^{-1}$, we reduce the degeneracy in $L^*_\mathrm{Sch}$ and $\alpha_\mathrm{Sch}$ and obtain $\alpha_\mathrm{Sch}$ by fitting the three Schechter parameters simultaneously. To study the evolution of the faint end slope of the Ly$\alpha$ LF, we compare our result with $\alpha_\mathrm{Sch}$ derived from previous photometric \citep{konno18,sobral18a} and spectroscopic \citep{drake17b,herenz19} studies. We fit a linear relationship between $\alpha_\mathrm{Sch}$ and $z$ to the measurements from this work and \citet{drake17b} in Figure \ref{fig:a_z}. Our best-fit linear function has a slope of $-0.38_{-0.26}^{+0.18}$ and an intersection of $-0.71_{-0.60}^{+0.86}$. To evaluate the strength of the linear relation between $\alpha_\mathrm{Sch}$ and $z$, we calculate the Pearson correlation coefficient $r$. We obtain the result of $r=-0.72$ with a $p$-value of $0.15$, indicating a marginally significant trend of anti-correlation between $\alpha_\mathrm{Sch}$ and $z$. Comparison between the $\alpha_\mathrm{Sch}-z$ relation of the Ly$\alpha$ LF and that for the galaxy UV LF \citep{parsa16} suggests that the $\alpha_\mathrm{Sch}$ of the Ly$\alpha$ LFs is steeper than that of the galaxy UV LFs. Also, the evolution of $\alpha_\mathrm{Sch}$ of the Ly$\alpha$ LF is more rapid than that of galaxy UV LF. The steepening of $\alpha_\mathrm{Sch}$ of the Ly$\alpha$ LFs towards higher redshift is consistent with the observational results that dust attenuation of Ly$\alpha$ emission from SF galaxies decreases towards fainter UV luminosity (e.g., \citealt{ando06,ouchi08}) and higher redshift (e.g., \citealt{blanc11,hayes11}). Because SF galaxies become less dusty at higher redshift and lower mass, a larger fraction of faint LAEs are observed towards higher redshift. The increase in the fraction of faint LAEs may contribute to the rapid increase in Ly$\alpha$ escape fraction from $z \sim 2$ to $z \sim 6$ \citep{konno16}.

\subsection{Type 1 AGN UV LF and the Evolution}\label{subsec:uvLF_evolution}
\begin{figure}[h!]
\begin{center}
\includegraphics[scale=0.4]{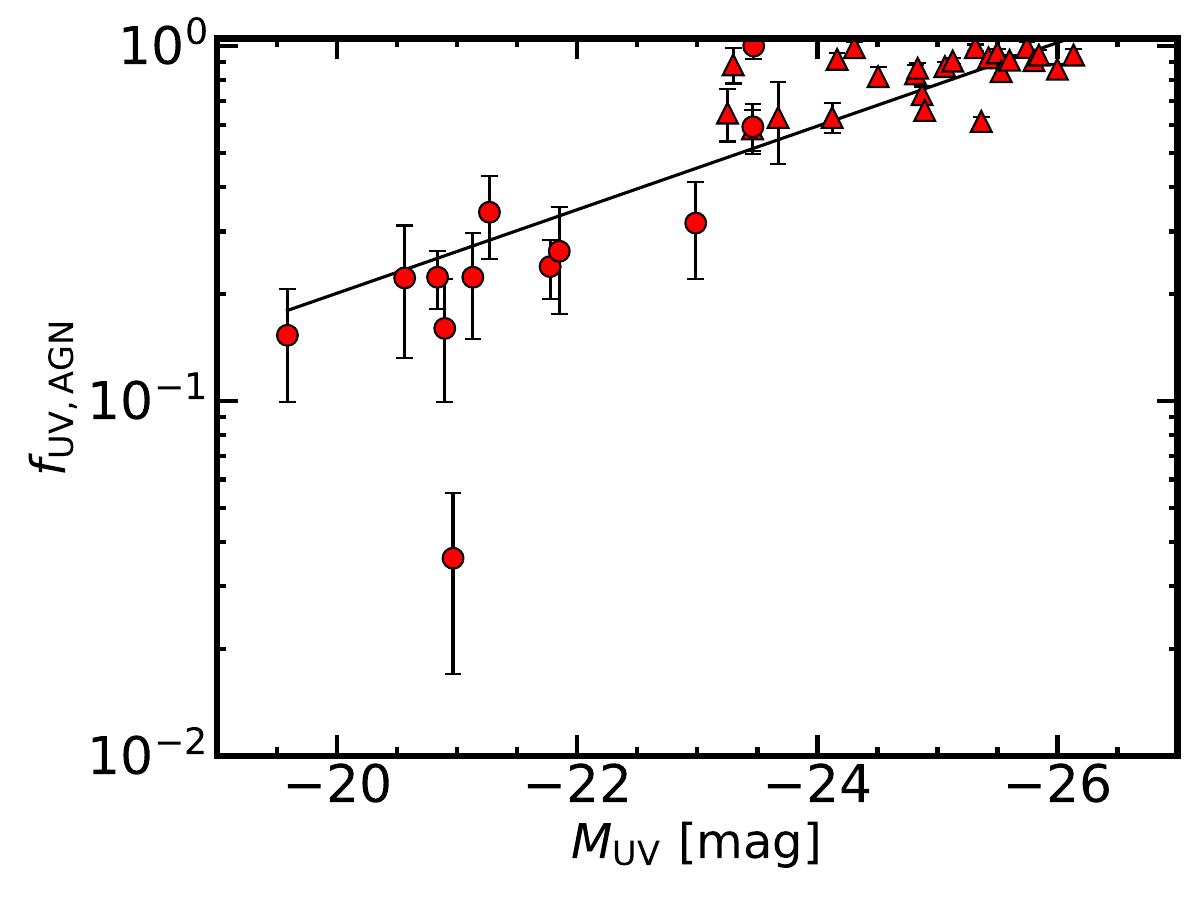}
\end{center}
\caption{$M_\mathrm{UV}-f_\mathrm{UV,AGN}$ relation derived by \citet{kakuma20}. The red circles (triangles) indicate type 1 AGN selected from HETDEX (SDSS). Our best-fit linear function of $\log f_\mathrm{UV,AGN} = -3.056 M_\mathrm{UV} -0.118$ is shown in the black solid line (see text).  \label{fig:fuv_agn}}
\end{figure}

\begin{figure}[ht!]
\begin{center}
\includegraphics[scale=0.4]{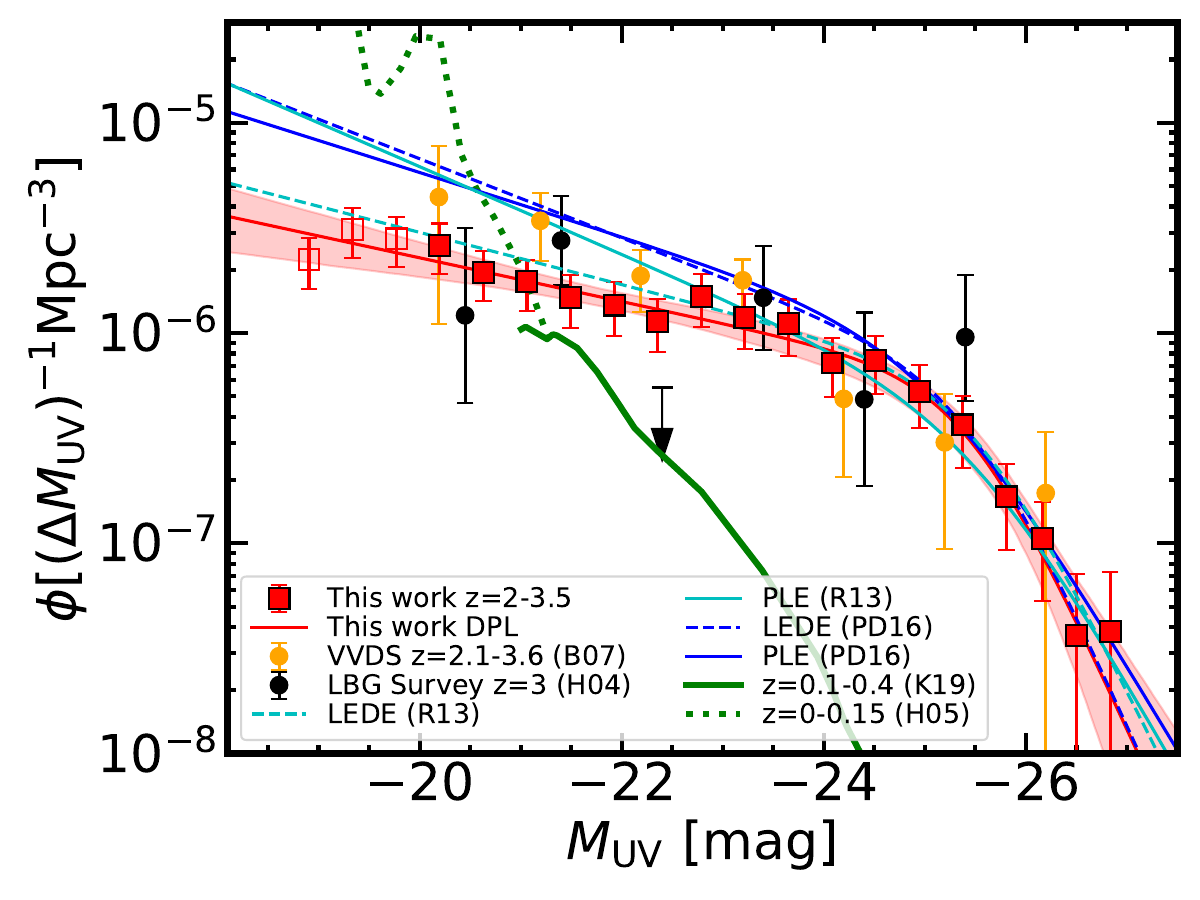}
\end{center}
\caption{Type 1 AGN UV LF. Our results at $z=2-3.5$ are presented in the red squares. Similar to Figure \ref{fig:LFuv}, data points that may be affected by incompleteness are marked with open red squares. The yellow circles denote the AGN UV LF from VVDS survey (\citealt{bon07}; B07) at $z=2.0-3.5$. The black circles indicate the AGN UV LF from LBG survey (\citealt{hunt04}; H04) at $z\sim 3$. The cyan and blue lines are the predicted QSO UV LF at $z=2.75$ from \citet[R13]{ross13} and \citet[PD16]{pd16}, respectively, based on the PLE (solid) and LEDE (dashed) models. The green dashed and solid lines represent the UF LF of type 1 Seyfert galaxies at $0 < z < 0.15$ (\citealt{hao05}, H05) and the type 1 AGN UV LF at $0.1 < z < 0.4$ (\citealt{kul19}, K19), respectively. \label{fig:LFuv_fit}}
\end{figure}

As mentioned in Section \ref{subsec:LF_uv}, our UV LFs of BL-LAEs are based on the total flux contributed from both the AGN and stellar components of host galaxies. Here we refer to the type 1 AGN UV LF as the UV LFs of BL-LAE with the flux contributed from the stellar components of galaxies removed. We derive the type 1 AGN UV LFs from UV LFs of BL-LAEs using the AGN UV flux ratio ($f_\mathrm{flux,AGN}$) that is defined as the ratio of flux contributed from AGN to the total flux. Specifically, we calculate $f_\mathrm{flux,AGN}$ of our BL-LAEs at each $M_\mathrm{UV}$ bin based on the results from \citet{kakuma20}. \citet{kakuma20} obtained a sample of 37 type 1 AGN at $z=2-3.5$ using the same method as mentioned in \ref{subsec:bllae} and conducted SED fitting on their sample with the Code Investigating GALaxy Emission ({\tt CIGALE}; \citealt{boq19}). Their results are shown in Figure \ref{fig:fuv_agn}. Assuming a linear relation between $M_\mathrm{UV}$ and $\log f_\mathrm{flux,AGN}$, we obtain the follwing best-fit model by MCMC:
\begin{equation} \label{eq:fuv_agn}
    \log f_\mathrm{UV,AGN} = -3.056 M_\mathrm{UV} -0.118
\end{equation}
We multiply the $M_\mathrm{UV}$ of the UV LF in Figure \ref{fig:LFuv} by $f_\mathrm{flux,AGN}$, obtaining the type 1 AGN UV LFs as shown in Figure \ref{fig:LFuv_fit}. Our results reach the very faint end of UV absolute magnitude at $M_\mathrm{UV} \sim\ -20$ and agree well with \citet{bon07}.  Following the procedure mentioned in Section \ref{subsec:LF_lya}, we fit our type 1 AGN UV LF with the DPL:
\begin{multline} \label{eq:mdpl}
    \phi_\mathrm{AGN}(M_\mathrm{UV}) \mathrm{d}M_\mathrm{UV} = \\
    \frac{\phi^*_\mathrm{AGN}\ \mathrm{d}M_\mathrm{UV}}{10^{(\alpha_\mathrm{AGN}+1)(M_\mathrm{UV}-M^*_\mathrm{AGN})} + 10^{(\beta_\mathrm{AGN}+1)(M_\mathrm{UV}-M^*_\mathrm{AGN})}},
\end{multline}
and obtain the best-fit parameters $\alpha_\mathrm{AGN}=-1.26^{+0.10}_{-0.08}$, $\beta_\mathrm{AGN}=-3.67^
{+0.80}_{-1.13}$, $M^*_\mathrm{AGN}=-25.44^{+0.50}_{-0.33}$, and $\log\phi^*_\mathrm{AGN}\mathrm{[Mpc^{-3}]}=-6.21^{+0.18}_{-0.16}$. Our best-fit AGN UV LF and the posterior distributions of the parameters are shown in Figure \ref{fig:LFuv_fit} and Figure \ref{fig:LFuv_fit_contour}, respectively. We compare our results with \citet{ross13} and \citet{pd16}, who studied the QSO UV LF using the pure luminosity evoution (PLE) model and the luminosity evolution and density evolution (LEDE) model. As presented in Figure \ref{fig:LFuv_fit}, our type 1 AGN UV LF is consistent with the LEDE model of \citet{ross13} at $z=2.75$, suggesting that the AGN UV LF evolve in both luminosities and number densities. We also compare our LFs at $z = 2-3$ with those at low redshifts, the type 1 AGN UV LF at $0.1<z<0.4$ \citep{kul19} and type 1 Seyfert UV LF at $0<z<0.15$ \citep{hao05}. In Figure \ref{fig:LFuv_fit}, a strong evolution of type 1 AGN UV LFs is identified. At the bright end with $M_\mathrm{UV}<-22$, we reproduce previous results that show the decrease of number density towards low redshift. At the faint end with $M_\mathrm{UV} > -22$, our results show that the number density increases towards low redshift. Such a decrease (increase) of bright (faint) AGN number density towards low redshift can be interpreted as the AGN downsizing effect. First identified in X-ray studies (e.g., \citealt{hasinger08,silverman08}), the AGN downsizing effect was observed in the UV LF only in the bright regime with $M_\mathrm{UV} \lesssim -21$ (e.g., \citealt{hopkins07,croom09,ikeda12}). Our results further support the AGN downsizing effect for the very faint AGN with the absolute UV magnitude down to $M_\mathrm{UV} \sim -20$.

The AGN downsizing effect may be connected with the quenching of massive galaxies that are observed in the local universe (e.g., \citealt{granato04,merloni04}). Massive BHs, which reside in massive host-galaxies, grow at a high accretion rate and make up the bright end of the AGN UV LFs at $z~\sim~2-3$. Towards lower redshift, the intense AGN activity in massive galaxies may gradually dispel gas from DM halos, resulting in the halt of star formation in the host galaxies as well as the decrease of BH accretion rate. The AGN become dimmer and populate the faint end of the UV LF. On the other hand, the moderate AGN feedback from less massive BHs would allow the host galaxies to keep their gas, supplying fuel to both star formation and BH accretion (e.g., \citealt{babic07,hirschmann14}).

To further investigate the redshift evolution of Ly$\alpha$ and UV LFs within in $z\sim 2-3$, a larger sample of LAE is needed. These analyses will be presented in our future studies that incorporate forthcoming HETDEX data release.

\section{Summary}\label{sec:summary}
We investigate the Ly$\alpha$ and UV LFs of LAEs at $z = 2-3.5$ that are obtained from the HETDEX spectroscopic survey. Our LAE sample include $16194$ SF galaxies and $2126$ type 1 AGN with EW$_0 > 20$~\AA~ that are spectroscopically identified in a sky area of $11.4$ deg$^2$. Our main results are summarized below.
\par
We derive the Ly$\alpha$ LF of LAEs at $z = 2-3.5$ in the Ly$\alpha$ luminosity range of $43.3 < \log L_\mathrm{Ly\alpha} [\mathrm{erg\ s^{-1}}] < \mathbf{45.5}$. Our Ly$\alpha$ LF is consistent with those from previous studies. At $\log L_\mathrm{Ly\alpha} [\mathrm{erg\ s^{-1}}] > 43.5$, our Ly$\alpha$ LF shows a clear bright-end hump. This is the first confirmation of such a bright-end hump with spectroscopically identified LAE samples. We confirm that the bright-end hump can be fully explained by type 1 AGN, which has been suggested in previous studies (e.g., \citealt{konno16,matthee17,sobral18a,calhau20,spinoso20}). 
\par
Combining our Ly$\alpha$ LF with the one derived by \citet{cassata11}, we show that the Ly$\alpha$ LF at $z\sim 2-3$ can be fitted with the combination of the Schechter function and the double power law. Our Ly$\alpha$ LF provides strong constraints at the bright end, reducing the degeneracy of the faint end slope$\alpha_\mathrm{Sch}$ and $L^*_\mathrm{Sch}$ by allowing all three Schechter parameters to be fitted simultaneously. From the Schechter component of our Ly$\alpha$ LF, we measure the faint end slope to be $\alpha_\mathrm{Sch} = -1.70^{+0.13}_{-0.14}$ at $z = 2-3.2$. We investigate the redshift evolution of $\alpha_\mathrm{Sch}$ based on our measurement and those from previous studies. We find that there is a possible redshift evolution of $\alpha_\mathrm{Sch}$ over $z \sim 2-6$ with a correlation coefficient $r=-0.72$ at the confidence level of $85\%$. We obtain a linear relation $\alpha_\mathrm{Sch} = -0.38_{-0.26}^{+0.18}\ z - 0.71_{-0.60}^{+0.86}$ at $z \sim 2-6$. Comparing with the linear relation of $\alpha_\mathrm{Sch}$ of the UV LF, we find $\alpha_\mathrm{Sch}$ of the Ly$\alpha$ LF may decrease more rapidly towards high redshift. 
\par
We derive the UV LF of LAEs at $z = 2-3.5$ in the range of $-27 < M_\mathrm{UV} < 20$, connecting the SF galaxy and AGN regimes. Our UV LF are consistent with previous results of QSO UV LF at $M_\mathrm{UV} < -23$ and LAE UV LF at $M_\mathrm{UV} > -21.5$. Combining our results and the UV LF of galaxies at $z~\sim~3$, we calculate the LAE number fraction as a function of $M_\mathrm{UV}$. Combining with previous measurements, we show that at $M_\mathrm{UV}$ fainter (brighter) than $M^*_\mathrm{UV} \sim -21$, $X_\mathrm{LAE}$ decreases (increases) with $M_\mathrm{UV}$.  
\par
We derive the type 1 AGN UV LF based on the UV LF of LAEs. Our type 1 AGN UV LF reaches a very faint absolute UV magnitude of $M_\mathrm{UV} \sim -20$ and agrees well with those from previous studies at similar redshift. Comparing with the results at lower redshifts, we find that the number density of faint ($M_\mathrm{UV} > -21$) AGN increases from $z \sim 2$ to $z \sim 0$. 
Such a number density evolution is compatible with AGN downsizing effect, which may have a close connection with the quenching of galaxy formation.


\acknowledgments
We thank the anonymous reviewer for the constructive comments that improve the quality of this work.
YZ thanks E. Komatsu for his help during the completion of this work.
SLF acknowledges support from the National Science Foundation through award AST-1908817.
CMC thanks the National Science Foundation for support through grants AST-1814034 and AST-2009577, the University of Texas at Austin College of Natural Sciences, and the Research Corporation for Science Advancement from a 2019 Cottrell Scholar Award sponsored by IF/THEN, an initiative of Lyda Hill Philanthropies. \par
HETDEX is led by the University of Texas at Austin McDonald Observatory and Department of Astronomy with participation from the Ludwig-Maximilians-Universit\"{a}t M\"{u}nchen, Max-Planck-Institut f\"{u}r Extraterrestrische Physik (MPE), Leibniz-Institut f\"{u}r Astrophysik Potsdam (AIP), Texas A\&M University, Pennsylvania State University, Institut f\"{u}r Astrophysik G\"{o}ttingen, The University of Oxford, Max-Planck-Institut f\"{u}r Astrophysik (MPA), The University of Tokyo and Missouri University of Science and Technology. In addition to Institutional support, HETDEX is funded by the National Science Foundation (grant AST-0926815), the State of Texas, the US Air Force (AFRL FA9451-04-2- 0355), and generous support from private individuals and foundations.
The observations were obtained with the Hobby-Eberly Telescope (HET), which is a joint project of the University of Texas at Austin, the Pennsylvania State University, Ludwig-Maximilians-Universit\"{a}t M\"{u}nchen, and Georg-August-Universit\"{a}t G\"{0}ttingen. The HET is named in honor of its principal benefactors, William P. Hobby and Robert E. Eberly.
VIRUS is a joint project of the University of Texas at Austin, Leibniz-Institut f\"{u}r Astrophysik Potsdam (AIP), Texas A\&M University, Max-Planck-Institut f\"{u}r Extraterrestrische Physik (MPE), Ludwig-Maximilians- Universit\"{a}t M\"{u}nchen, The University of Oxford, Pennsylvania State University, Institut f\"{u}r Astrophysik G\"{o}ttingen,,  Max-Planck-Institut f\"{u}r Astrophysik (MPA)
The Texas Advanced Computing Center (TACC) at The University of Texas at Austin provided high performance computing, visualization, and storage resources that contributed to the research results reported within this paper. \par
The Institute for Gravitation and the Cosmos is supported by the Eberly College of
Science and the Office of the Senior Vice President for Research at The
Pennsylvania State University. \par
The Hyper Suprime-Cam (HSC) collaboration includes the astronomical communities of Japan and Taiwan, and Princeton University. The HSC instrumentation and software were developed by the National Astronomical Observatory of Japan (NAOJ), the Kavli Institute for the Physics and Mathematics of the Universe (Kavli IPMU), the University of Tokyo, the High Energy Accelerator Research Organization (KEK), the Academia Sinica Institute for Astronomy and Astrophysics in Taiwan (ASIAA), and Princeton University. Funding was contributed by the FIRST program from the Japanese Cabinet Office, the Ministry of Education, Culture, Sports, Science and Technology (MEXT), the Japan Society for the Promotion of Science (JSPS), Japan Science and Technology Agency (JST), the Toray Science Foundation, NAOJ, Kavli IPMU, KEK, ASIAA, and Princeton University. 

This paper makes use of software developed for the Large Synoptic Survey Telescope. We thank the LSST Project for making their code available as free software at \url{http://dm.lsst.org}.

This research is based in part on data collected at Subaru Telescope, which is operated by the National Astronomical Observatory of Japan. We are honored and grateful for the opportunity of observing the Universe from Maunakea, which has the cultural, historical and natural significance in Hawaii. \par
This paper is supported by World Premier International Research Center Initiative (WPI Initiative),
MEXT, Japan, and KAKENHI Grant-in-Aid for Scientific Research (A)
(19H00697, 20H00180, and 21H04467) through the Japan
Society for the Promotion of Science (JSPS). 


\clearpage
\appendix

\begin{figure*}[hb!]
\begin{center}
\includegraphics[scale=0.45]{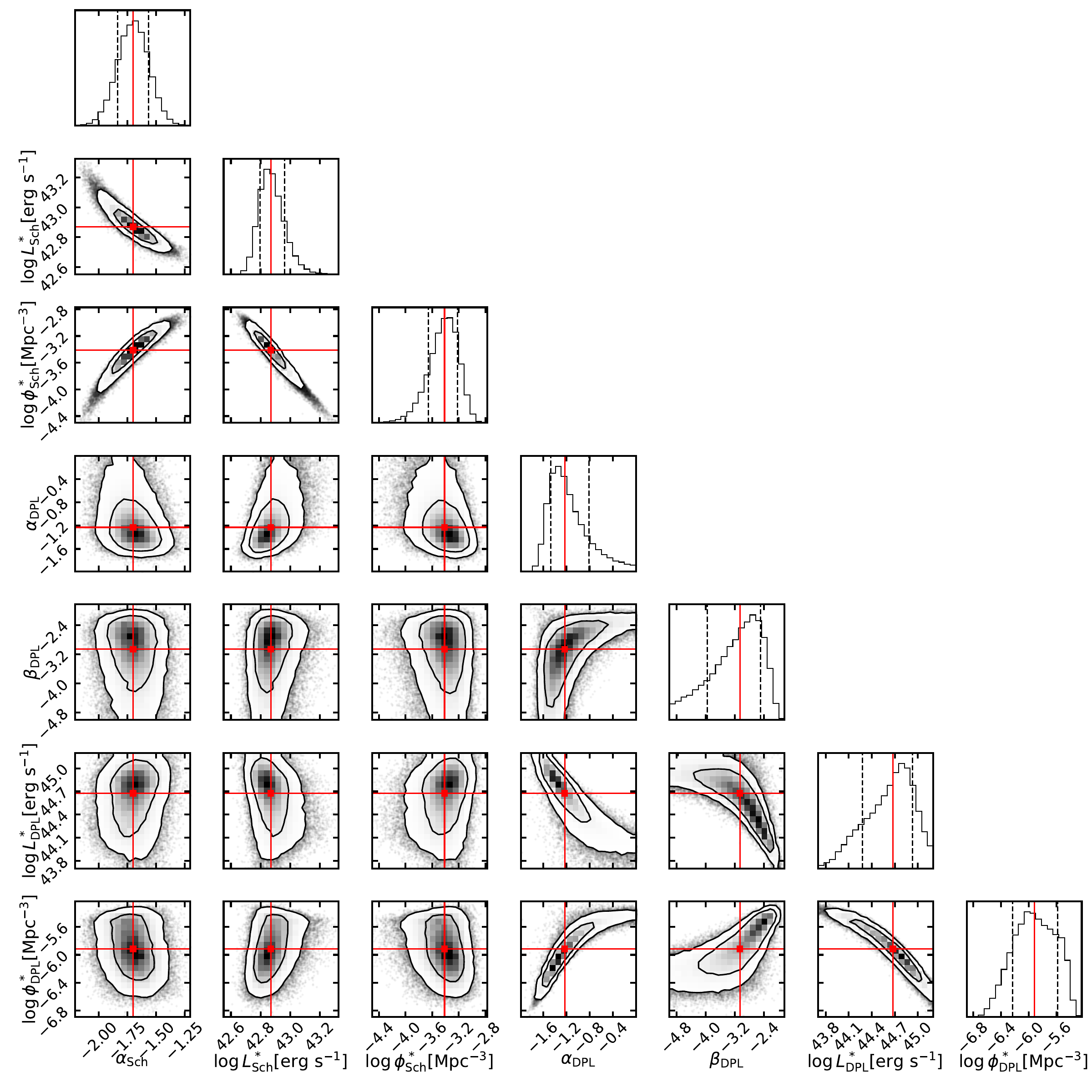}
\end{center}
\caption{Error contours of our best-fit Ly$\alpha$ LF (Model 1). The red crosses denotes the best-fit results of the parameters. The inner (outer) contour indicates the $68.3\%$ ($95.4\%$) confidence level. \label{fig:LF_fit_contour}}
\end{figure*}

\begin{figure*}[ht!]
\begin{center}
\includegraphics[scale=0.5]{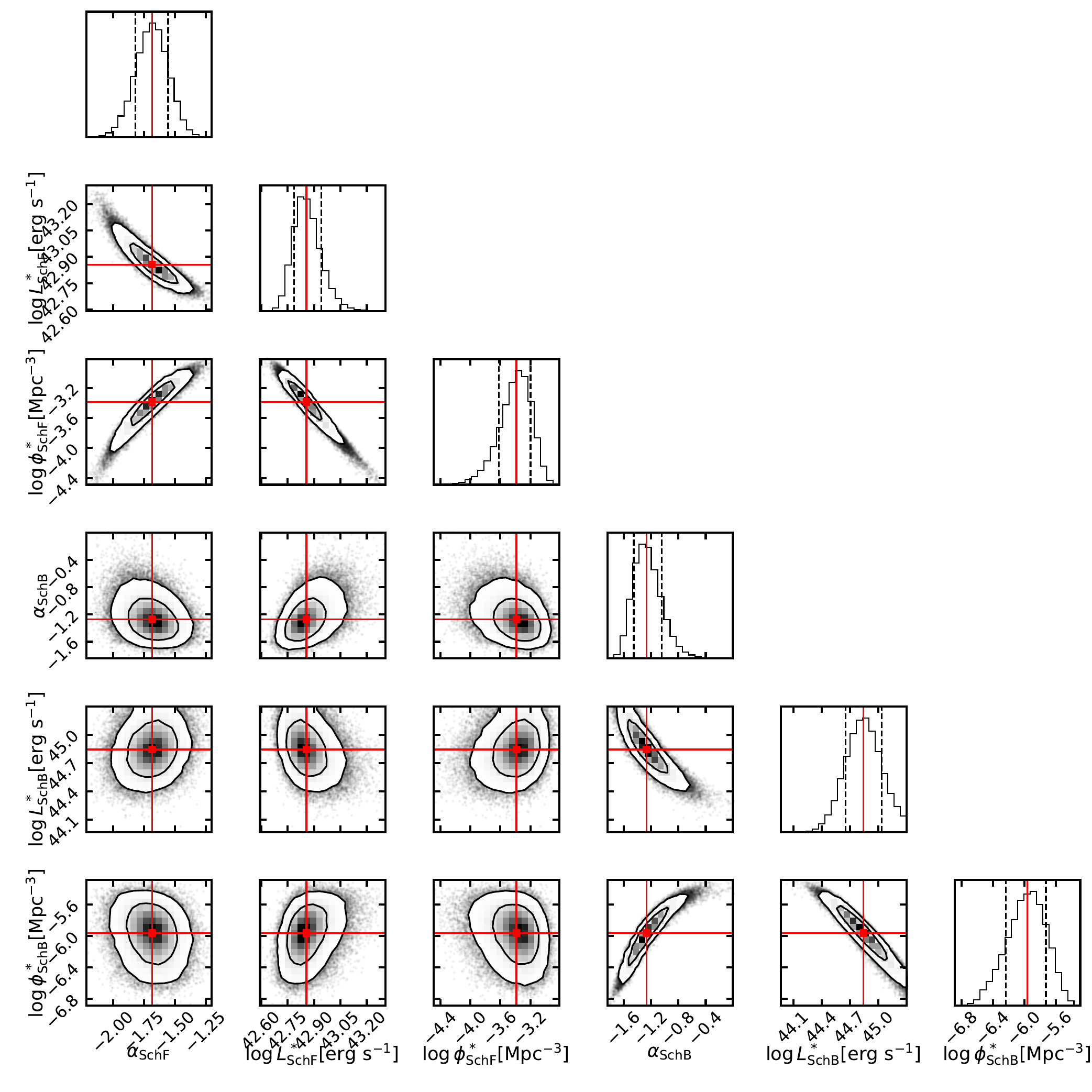}
\end{center}
\caption{Same as Figure \ref{fig:LF_fit_contour}, but for Model 2. The Schechter parameters of the faint and bright component are denoted with subscriptions of ``SchF" and ``SchB", respectively. \label{fig:LF_fit_contour_m2}}
\end{figure*}

\begin{figure*}[ht!]
\begin{center}
\includegraphics[scale=0.6]{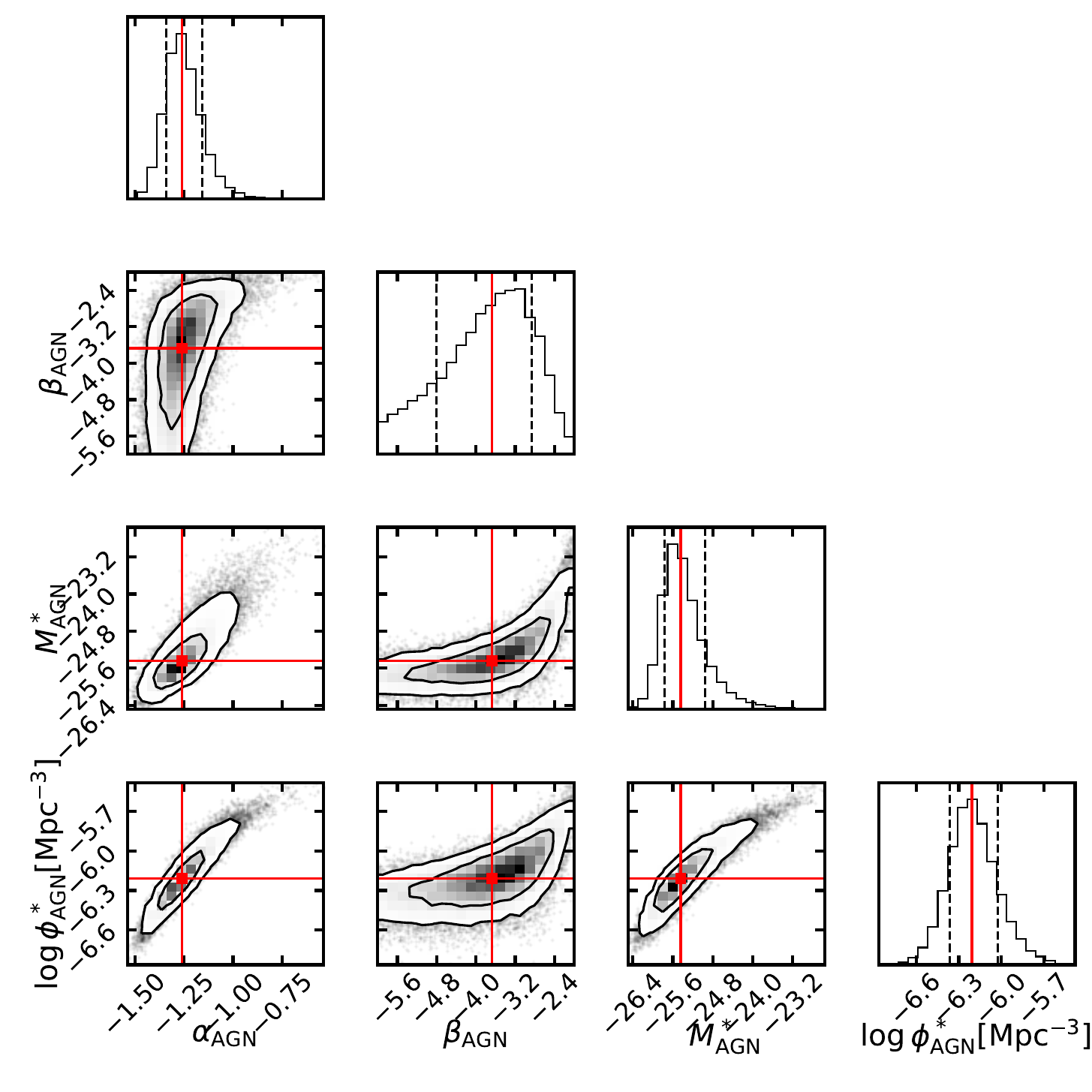}
\end{center}
\caption{Same as Figure \ref{fig:LF_fit_contour}, but for type 1 AGN UV LF. \label{fig:LFuv_fit_contour}}
\end{figure*}

\end{document}